\documentclass[letterpaper, 10 pt, conference]{ieeeconf}  

\usepackage{multicol}
\usepackage[bookmarks=true]{hyperref}
\usepackage{graphicx}
\usepackage{amsmath}
\usepackage{xcolor}
\usepackage{algorithm}
\usepackage{algpseudocode}
\usepackage{amsmath}
\usepackage{booktabs}
\usepackage{siunitx}
\usepackage{tikz}
\usepackage{booktabs}
\usepackage{algorithm}
\usepackage{algpseudocode}
\usepackage[T1]{fontenc}

\newcommand{\addon}[1]{\textcolor{blue}{#1}}

\DeclareMathOperator{\LP}{LPFilter}

\newcommand{\X}{\mathcal{X}}
\newcommand{\U}{\mathcal{U}}

\newcommand{\cost}{c}
\newcommand{\costf}{\cost_{f}}

\newcommand{\cutoff}{f_{c}}
\newcommand{\order}{o_{\text{LPF}}}

\IEEEoverridecommandlockouts                              

\title{\LARGE \bf
LP-MPPI: Low-Pass Filtering for Efficient \\Model Predictive Path Integral Control
}

\author{Piotr Kicki
\thanks{Piotr Kicki is with Institute of Robotics and Machine Intelligence, Poznan University of Technology, Poznan, Poland and with IDEAS Research Institute, Warsaw, Poland {\tt\small piotr.kicki@put.poznan.pl}}%
\thanks{This work was supported by the National Science Centre, Poland, under the INTENTION project (grant no. 2021/43/I/ST6/02711) and by the infrastructure of the Poznan Supercomputing and Networking Center.}%
}

\begin{document}

\maketitle
\thispagestyle{empty}
\pagestyle{empty}

\begin{abstract}
Model Predictive Path Integral (MPPI) control is a widely used sampling-based approach for real-time control, valued for its flexibility in handling arbitrary dynamics and cost functions. However, it often suffers from high-frequency noise in the sampled control trajectories, which hinders the search for optimal controls and transfers to the applied controls, leading to actuator wear. 
In this work, we introduce Low-Pass Model Predictive Path Integral Control (LP-MPPI), which integrates low-pass filtering into the sampling process to eliminate detrimental high-frequency components and enhance the algorithm's efficiency. Unlike prior approaches, LP-MPPI provides direct and interpretable control over the frequency spectrum of sampled control trajectory perturbations, leading to more efficient sampling and smoother control. Through extensive evaluations in Gymnasium environments, simulated quadruped locomotion, and real-world F1TENTH autonomous racing, we demonstrate that LP-MPPI consistently outperforms state-of-the-art MPPI variants, achieving significant performance improvements while reducing control signal chattering. 
\end{abstract}

\section{INTRODUCTION}

One of the key abilities of the autonomous system is to determine the best actions given a certain goal, i.e. real-time motion planning and control. 
If the model of the controlled system is available, one of the best performing approaches is Model Predictive Control (MPC), which has proven its capabilities to solve many challenging tasks, such as autonomous racing~\cite{racingmpc}, off-road driving~\cite{mppi}, agile drone flight~\cite{krinner_mpcc_2024}, and legged locomotion~\cite{leggedmpc}.

In general, there are two main approaches to MPC: sampling-based and optimization-based. 
Optimization-based MPC algorithms provide an efficient way to find optimal control sequences using dynamics and cost function gradients~\cite{sqprti, acados}.
However, they typically impose requirements on the dynamics model or cost function formulations, such as differentiability or continuity.
An interesting alternative is sampling-based MPC. One of the main benefits of this approach is that the dynamics and cost functions can be arbitrary, and the only requirement is to evaluate them relatively fast. 
The two algorithms within this group, which have proven their effectiveness in numerous tasks, such as off-road driving~\cite{mppi}, drone flight~\cite{dronemppi}, and control of high-dimensional simulated systems (humanoids, dexterous hands, manipulators)~\cite{icem, storm2021}, are the Cross Entropy Method (CEM)~\cite{cem} and Model Predictive Path Integral Control (MPPI)~\cite{mppi}.

\begin{figure}[t]
    \centering
    \includegraphics[width=\linewidth]{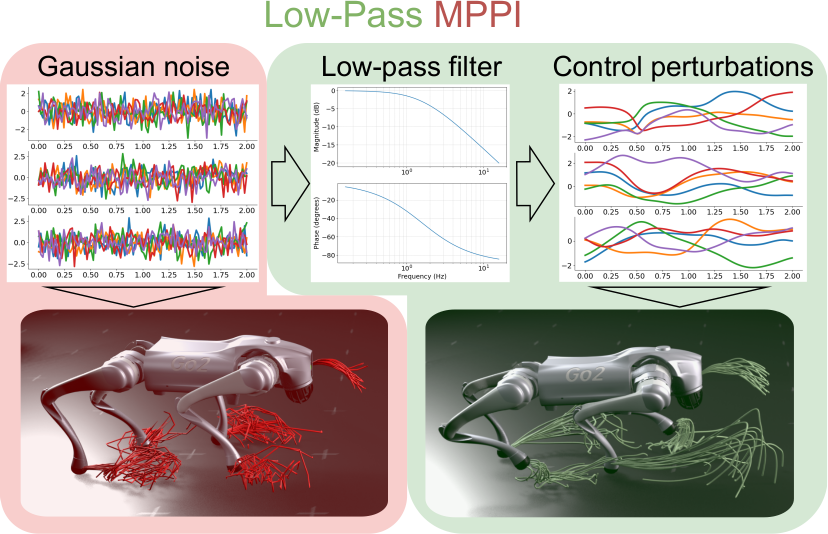}
    \vspace{-5mm}
    \caption{The proposed Low-Pass Model Predictive Path Integral (LP-MPPI) control smooths out the controls via the low-pass filtering of control perturbations and increases the efficiency of the MPPI.}
    \label{fig:cover}
\end{figure}

The core of both MPPI and CEM approaches is to evaluate the performance of the control trajectories sampled from a sequence of Gaussian distributions. 
This approach results in a potential lack of temporal correlation within individual trajectories (see the red part of Figure~\ref{fig:cover}). In fact, control trajectory perturbations are approximately white noise signals, i.e., they are evenly composed of all possible frequencies. This, in turn, results in the overrepresentation of the high-frequency components when compared to the expected optimal behaviors in most robotic systems. 
Moreover, the dynamics of most robots naturally dampen these components, so their impact on the resulting performance is minimal.
Thus, they do not contribute to the efficient search for optimal control signals but instead result in chattering of the applied controls and wear out of the actuators.


To address these issues, researchers proposed several interesting approaches, such as spline interpolation of control signals~\cite{storm2021, splinemppi} or input-lifting~\cite{smppi}. However, they offer only an indirect and coarse control over the control signal frequency spectrum. A more direct approach, inspired by the iCEM algorithm~\cite{icem}, was proposed in~\cite{colored}, where colored noise was used in MPPI instead of the white one. However, colored noise damps the signal components proportionally to the inverse of their frequency, or its powers, which may not be desired, especially for tasks that require frequent repetitiveness, e.g., legged locomotion, stirring, or chopping.

In this paper, we address these issues by extending the MPPI algorithm with a low-pass filter applied to the sampled control perturbations. Our method, called Low-Pass Model Predictive Path Integral control (LP-MPPI), is presented schematically in Figure~\ref{fig:cover}.
Our approach eliminates the harmful and ineffective high-frequency components of the control signal perturbations, increasing the efficiency of the MPPI algorithm and smoothing the resultant control signal without sacrificing its responsiveness. Moreover, unlike colored noise, it does not bias the control signal frequencies in the filter's passband, enabling an effective search for high-performing trajectories across the admissible frequency range. 
Our method introduces only two interpretable parameters with clear physical meaning, allowing intuitive tuning. 
They directly control the frequency range in which the search for the optimal control trajectory is focused and the damping characteristics beyond it.
Finally, LP-MPPI is easy to implement and adds negligible computational overhead compared to the MPPI algorithm.


We conduct an extensive experimental evaluation of the proposed method against the State-of-the-Art variants of the MPPI in three Gymnasium environments, over a wide range of MPPI parameters. The results demonstrate that our approach consistently outperforms the considered baselines, improving their results by 24\% on average, while significantly reducing chattering in the applied control signals.
To further assess its applicability, we integrate LP-MPPI with the recently developed Dial-MPC~\cite{dialmpc} and evaluate the resulting LP-Dial-MPC on quadruped robots. This experiment highlights the ease and effectiveness with which our method can be combined with other MPPI-based approaches, yielding an average performance gain of over 32\% compared to Dial-MPC.
Finally, we validate our approach in real-world autonomous F1TENTH racing, where it outperforms all baselines, most of them by a large margin.

Our contributions can be summarized as follows:
\begin{itemize}
    \item We propose a Low-Pass Model Predictive Path Integral algorithm, which enables shaping the frequency spectrum of the control trajectories' perturbations distribution, improving search efficiency and reducing the high-frequency noise in the applied control signal.
    \item We conduct a thorough experimental analysis of the proposed approach in several simulated environments, showing that the proposed method consistently outperforms the state-of-the-art MPPI-based control approaches, while reducing the amount of high-frequency components in the applied control signals (see Tab.~\ref{tab:smoothness}).
    \item We highlight the practicality of the proposed solution, that is, the ease of implementation of our method and its integration with the latest sampling-based MPC methods, the physical interpretability of its parameters, and the intuitiveness of tuning them.
\end{itemize}


\section{Related Work}

In recent years, many attempts have been made to improve the effectiveness of the MPPI algorithm~\cite{storm2021, splinemppi, smppi, colored, dialmpc}. A substantial number of these methods considered how to increase the smoothness of the control signal and improve the sampling efficiency.

In~\cite{smppi}, authors proposed to smooth the control signals generated by MPPI, by (i) lifting the input-space, i.e., sampling in the space of the derivatives of the actual controls, and (ii) adding a smoothness cost. This approach effectively transforms the frequency spectrum of the control signals' perturbations into colored noise and gives only minimal control of the frequency spectrum.
Our frequency-based shaping of control signal perturbations can adjust the time correlation of the trajectories in a more fine-grained way, does not introduce such a strong bias to the lowest frequencies, and does not require additional cost terms. 
Moreover, it should be noted that adding smoothness costs reduces search efficiency because part of the drawn trajectories will be effectively excluded due to high smoothness costs.

Interestingly, the authors of~\cite{smppi} showed that using a low-pass Savitzky-Golay filter on the control trajectory update signal (see MPPI(SGF) in~\cite{smppi}) results in a poorer performance than their proposed SMPPI. Our work shows that if the low-pass filtering is applied before the control perturbations are simulated, it maintains constraint satisfaction, introduces no phase distortion, and outperforms SMPPI~\cite{smppi}.

Another interesting strategy to smooth the control signals and improve exploration efficiency is to exploit spline interpolation~\cite{storm2021, splinemppi, howell2022predictivesampling}. Instead of sampling a long sequence of actions, one can sample only a reduced number of spline control points and then interpolate between them. Thus, the dimension of the exploration space is reduced, and the resulting trajectories are smoother. However, in this approach, the control over the smoothness is indirect and rather coarse and may require predicting the time increments along the controls~\cite{howell2022predictivesampling, kicki2025corl}.

A more direct approach to improving control smoothness and exploration capabilities is to exploit some specific distributions from which the control perturbation samples are drawn, as the default Gaussian one may not be an optimal choice. In this theme, the authors of \cite{logmppi} propose to use a mixture of the original normal distribution with the log-normal distribution to increase the probability of sampling the larger deviations from the nominal control sequence to improve the exploration. In turn, in~\cite{gmmmppi} and \cite{svmppi}, authors explored the use of multimodal distributions, i.e., Gaussian Mixture Models and empirical particle-based distributions with Stein Variational Gradient Descent updates, to increase the exploration capabilities by allowing one to maintain multiple hypotheses. Recently, learning-based approaches have also been used to determine better sampling distributions based on the structure of the environment~\cite{learningmppi1, learningmppi2}, enabling focusing the search on promising areas of the state space.
Nevertheless, none of these methods considers the correlation between subsequent controls within each sampled control trajectory perturbation, which may result in a chattering of the control signal.

Methods that adjust the sampling distribution to address this particular issue, while increasing the exploration capabilities of the sampling-based MPC, were proposed in \cite{icem} and \cite{colored}. The approach presented in~\cite{icem}, called iCEM, is an improved version of the Cross-Entropy Method (CEM)~\cite{cem}. It shows that sampling the actions from the colored-noise distributions significantly improves the original CEM approach. The authors of~\cite{colored} employed the same idea in the MPPI algorithm, demonstrating improved smoothness and performance. In our work, we also focus on shaping the MPPI sampling distribution in the frequency domain. However, instead of relying on colored noise, we use a low-pass filter, which provides greater flexibility and enables uniform sampling across the desired range of low frequencies.

\section{Method}
\subsection{Problem definition}
We consider the following optimal control problem
\resizebox{\columnwidth}{!}{
\begin{minipage}{\columnwidth}
\begin{align}
\label{eq:ocp}
    \min_{u_{t:t+H}} & J(x_{t:t+H+1}, u_{t:t+H}) = \sum_{h=0}^{H} \cost(x_{t+h}, u_{t+h}) + \costf(x_{t+H+1})\nonumber\\
    & \text{subject to}\\
    & x_{t+h+1} = f(x_{t+h}, u_{t+h}), \quad \forall h \in \{0, ..., H\}\nonumber\\
    & x_{t:t+H+1} \in \X, u_{t:t+H} \in \U, \nonumber
\end{align}
\vspace{-2mm}
\end{minipage}
}
where $x_t$ and $u_t$ are the state and control signal at time $t$, $f$ is the system dynamics, $\cost$ and $\costf$ are step and terminal cost functions, $J$ is the total cost of the trajectory, while $\X$ and $\U$ are the sets of admissible states and controls, respectively.

\subsection{Model Predictive Path Integral Control}
Our proposed method, Low-Pass Model Predictive Path Integral Control (LP-MPPI), is based strongly on the original MPPI algorithm~\cite{mppi}. Therefore, we recall its pseudocode in Algorithm~\ref{alg:mppi}. In general, the idea is to (i) draw multiple control trajectories around the current nominal control trajectory, (ii) simulate them using the model of the system, (iii) compute their costs, and finally (iv) update the nominal trajectory based on the costs obtained by the perturbed controls. In this paper, we focus on the commonly overlooked aspect of the MPPI algorithm -- drawing random control perturbations. In fact, the only part of the algorithm that we would like to analyze and improve is located in line \ref{lst:line:noise}.
To do so, we will analyze the original MPPI algorithm~\cite{mppi} and its recent extension~\cite{colored} from a frequency domain perspective.

\begin{algorithm}
\caption{Model Path Integral Control (MPPI)}
\label{alg:mppi}
\begin{algorithmic}[1]
\Require Nominal control sequence $U$, current state $x_t$, System dynamics model $f$, step and terminal cost functions $\cost$, $\costf$, number of rollouts $N$, control sequence horizon $H$, temperature parameter $\lambda$, noise covariance matrix $\Sigma$
\Ensure Updated nominal control sequence $U$

\For{$i = 1$ to $N$}  \Comment{Sample $N$ trajectories}
    \State Sample noise sequence $\epsilon_{i,1:H}$ from $\mathcal{N}(0, \Sigma)$ \label{lst:line:noise}
    \State Generate control sequence $U_i = U + \epsilon_i$
    \State Compute trajectory $x_{i,t:t+H+1}$ using system dynamics $x_{i,t+h+1} = f(x_{i,t+h}, u_{i,h})$
    \State Compute cost $J_i = J(x_{i,t:t+H+1}, u_{i,1:H})$
\EndFor

\State Compute importance weights $w_i = \frac{e^{-\lambda J_i}}{\sum_{j=1}^{N} e^{-\lambda J_j}}$

\State Update controls $U = \sum_{i=1}^{N} w_i U_i$

\State Apply first control input $u_1$ to the system
\State Shift control sequence: $U \leftarrow \{u_2, ..., u_H, u_{H+1}\}$
\end{algorithmic}
\end{algorithm}

\subsection{Spectral analysis}
In the original MPPI, all elements of the noise sequences are sampled from the Gaussian distribution with some constant covariance matrix $\Sigma$. This results in a lack of correlation between the subsequent elements of the noise sequence. In general, this may be seen as an advantage, as perturbing the system with white noise is a well-known technique in system identification~\cite{sysid}, which ensures that the system dynamics are excited with all possible control frequencies. In the context of MPC, similar implications should also be true for the cost function $J$, which depends on both controls and state trajectories. However, in the context of robotic applications, it is pretty uncommon to control systems that require an excitation with high-frequency noise, as most robots filter it out. Therefore, control trajectories with high-frequency noise and low-frequency noise result in similar costs, which may result in frequent changes of the following actions, and thus jittering in the real system. Moreover, we expect that for most robotic systems, the spectrum of frequencies of the optimal control trajectories is not uniformly distributed, so the use of white noise may reduce the search efficiency and result in a performance decrease. In fact, using the uncorrelated noise distributions corresponds to the maximum exploration in the frequency spectrum.

To limit the amount of high-frequency noise in the sampled controls, the authors of~\cite{colored} proposed biasing the spectrum of the control trajectory perturbations towards low frequencies. As a result, they obtained a strategy that puts most of the signal energy in the lowest possible frequencies and gradually reduces the energy of the higher-frequency components.
In turn, in this paper, we analyze an alternative approach based on low-pass filtered noise, which balances the bias toward lower frequencies with efficient exploration of the admissible spectrum, and enables explicit control of this trade-off through adjustable parameters -- cutoff frequency and filter order.

\begin{figure}[t]
    \centering
    \setlength\tabcolsep{0.5mm}
    \begin{tabular}{cc}
         \scriptsize \textbf{Ant-v3} & \scriptsize \textbf{Humanoid-v3} \\
         \includegraphics[height=29mm]{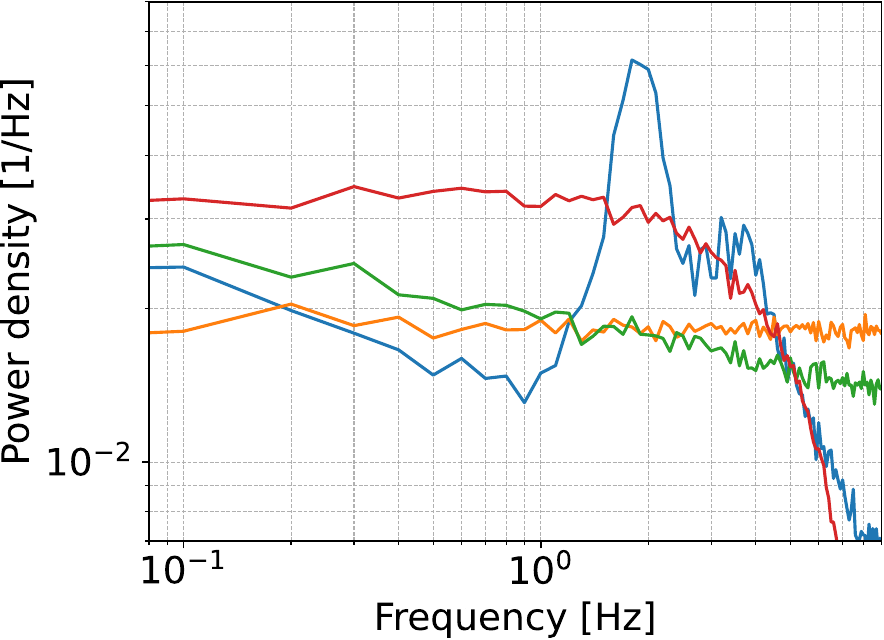} &
         \includegraphics[height=29mm]{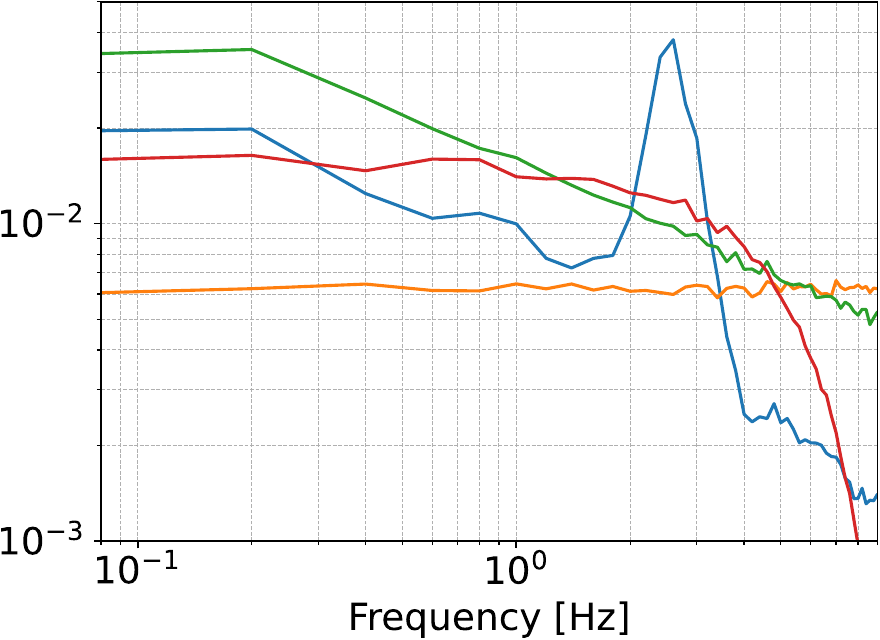}\\
    \end{tabular}
    \includegraphics[width=0.95\linewidth]{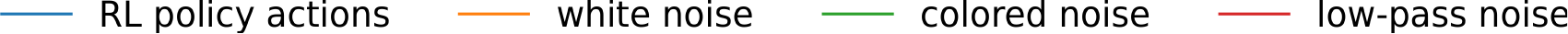}
    \vspace{-1mm}
    \caption{Spectrograms of the actions drawn from the trained RL policy and the different sampling distributions. Both white and colored noise are unable to closely fit the spectrum of RL behaviors, while the low-pass filtered noise covers it quite accurately.}
    \label{fig:spectral}
    \vspace{-3mm}
\end{figure}
To analyze each of the aforementioned control sampling strategies and provide the rationale for the approach proposed in this paper, we conducted an experiment. We compared the power density spectra of (i) white noise, (ii) colored noise, and (iii) low-pass filtered white noise, to the spectrum of the control signals generated by a trained RL policy from the StableBaselines3-zoo library~\cite{rlzoo3} for two sample environments, Ant-v3 and Humanoid-v3. We chose the RL agent controls spectrum as a reference, as it may be considered to be very close to the frequency statistics of the optimal behaviors in these environments. 
To present the results in a compact form, we averaged the spectra of the individual joints for the RL agent.
In addition, we averaged the spectra over 100 seeds to reduce noise. Moreover, to ensure a fair comparison between the analyzed sampling distributions, we optimized their parameters to minimize the norm between their and the RL agent action spectra.
The results of this comparison are presented in Figure~\ref{fig:spectral}. 
One can see that in the considered locomotion tasks, the RL agent does not perform high-frequency actions; however, following only the lowest frequencies is also far from optimal, as it ignores the frequency bumps that occur around \SI{2}{\hertz}. In fact, the spectrum that best matches the RL agent is the low-pass-filtered white noise, as it does not damp the signal at the most important frequencies. Moreover, what is also important in the context of MPPI, it allows for efficient exploration in the assumed bandwidth.

\subsection{Low-pass Model Predictive Path Integral Control}
Following the observations made in the previous section, we introduce the Low-pass Model Predictive Path Integral Control (LP-MPPI). 
The core of the proposed algorithm is the introduction of a temporal correlation between the subsequent control perturbations by using a low-pass filter on them.
The proposed approach is formalized in Algorithm~\ref{alg:lpmppi}, where the changes with respect to the original MPPI algorithm are marked in blue.
Our method introduces a subtle yet important modification that (i) biases the search for optimal control trajectories toward low-frequency signals and (ii) allows the user to control the frequency spectrum within which the algorithm searches for candidate trajectory updates.
These features can be precisely controlled by tuning the parameters of the low-pass filter, which we describe in detail in Section~\ref{sec:parameters}. Importantly, the proposed modifications do not introduce significant computational overhead to the MPPI algorithm, as filtering typically requires a small number of multiplications proportional to the horizon length $H$ and, in most cases, is significantly cheaper than evaluating the system's dynamics.

One of the crucial design choices is to decide which signals should be smoothed in order to improve efficiency and reduce chattering without sacrificing controller reactiveness or violating system constraints. In the proposed approach, we apply low-pass filtering to sampled noise sequences, i.e. control trajectory perturbations (see line 2 of Algorithm~\ref{alg:lpmppi}). 
This way, we strongly bias the search for control trajectories towards low frequencies, but we do not constrain the applied control signal frequency spectrum. The first control may change rapidly, but the change of the entire control trajectory is correlated. Thus, the search is biased, but the applied control signal maintains its reactiveness. Moreover, unlike existing filtering approaches, such as MPPI(SGF)~\cite{smppi}, our approach improves search efficiency by evaluating only smoothed constraint-compliant perturbations and does not introduce any phase shift to the applied control signal, as if the filtering were applied directly to it.

In our software implementation, we utilized a simple and computationally efficient digital implementation of the Butterworth filter~\cite{butter}. This type of filter is maximally flat in the passband, which prevents introducing biases to specific frequencies within the desired range. In turn, its frequency spectrum is not as steep as, for example, that of Chebyshev filters; however, as can be seen in Figure~\ref{fig:spectral}, this property is not necessarily needed in the context of sampling performant control trajectories. Similarly, the relatively high phase shift introduced by the Butterworth filter is not problematic in our approach, since temporally uncorrelated perturbations remain unaffected by such shifts.

\begin{algorithm}[t]
\caption{Low-pass Model Path Integral Control (LP-MPPI)}
\label{alg:lpmppi}
\begin{algorithmic}[1]
\Require Nominal control sequence $U$, current state $x_t$, System dynamics model $f$, step and terminal cost functions $\cost$, $\costf$, number of rollouts $N$, control sequence horizon $H$, temperature parameter $\lambda$, noise covariance matrix $\Sigma$, \addon{low-pass filter cutoff frequency $\cutoff$ and order $\order$}
\Ensure Updated nominal control sequence $U$

\For{$i = 1$ to $N$}  \Comment{Sample $N$ trajectories}
    \State Sample noise sequence $\epsilon_{i,1:H}$ from $\mathcal{N}(0, \Sigma)$ 
    \State \addon{Filter noise sequence $\epsilon^{\text{LP}}_{i} = \LP(\epsilon_{i}, \cutoff, \order)$ using low-pass filter}\label{lst:line:filter}
    \State Generate control sequence $U_i = U + \addon{\epsilon^{\text{LP}}_{i}}$
    \State Compute trajectory $x_{i,t:t+H+1}$ using system dynamics $x_{i,t+h+1} = f(x_{i,t+h}, u_{i,h})$
    \State Compute cost $J_i = J(x_{i,t:t+H+1}, u_{i,1:H})$
\EndFor

\State Compute importance weights $w_i = \frac{e^{-\lambda J_i}}{\sum_{j=1}^{N} e^{-\lambda J_j}}$

\State Update controls $U = \sum_{i=1}^{N} w_i U_i$

\State Apply first control input $u_1$ to the system
\State Shift control sequence: $U \leftarrow \{u_2, ..., u_H, u_{H+1}\}$
\end{algorithmic}
\end{algorithm}

\subsection{Parameters of LP-MPPI}
\label{sec:parameters}
\begin{figure}[t!]
        \newcommand{\bh}{31mm}
        \centering
        \includegraphics[height=\bh]{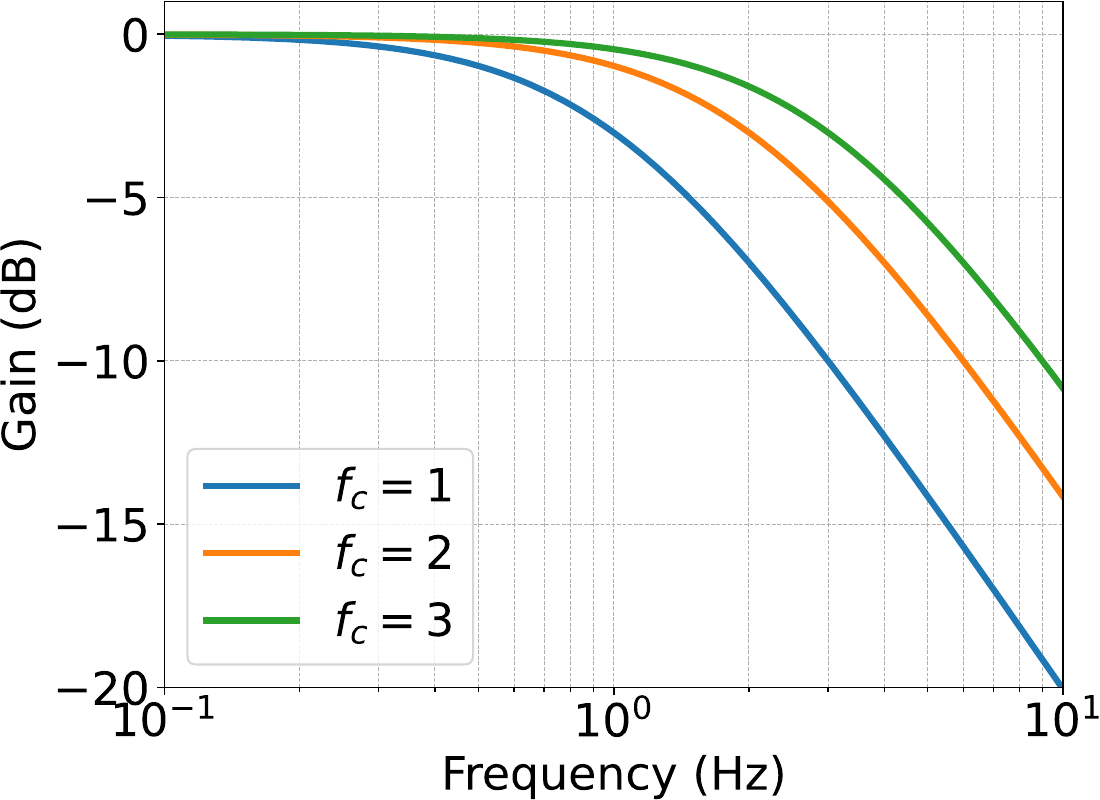}
        \includegraphics[height=\bh]{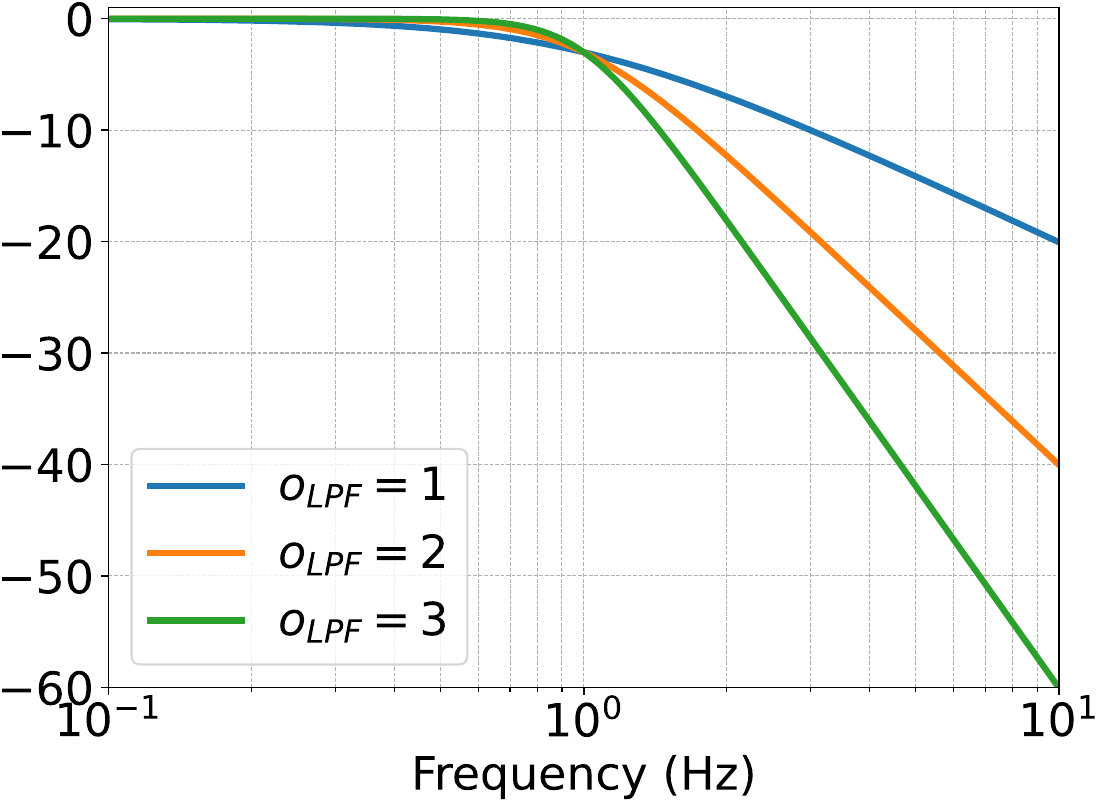}\\
        \vspace{3mm}
        \includegraphics[width=\linewidth]{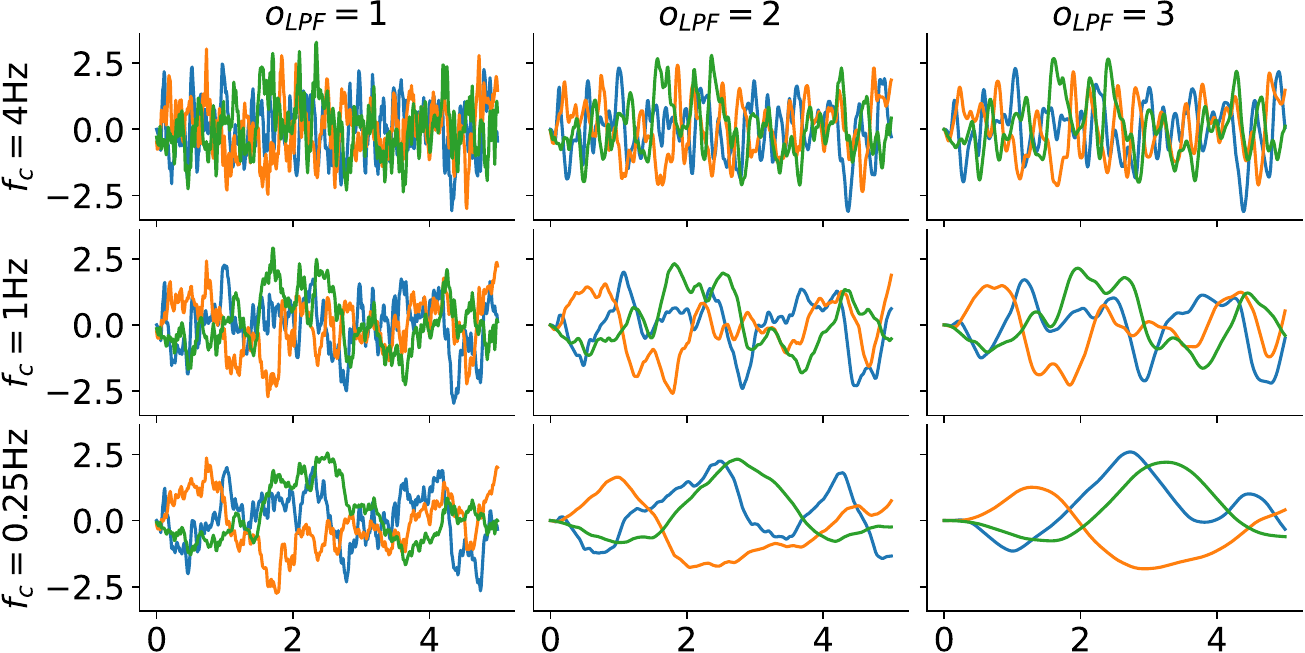}
    \vspace{-5mm}
    \caption{Frequency response of the Butterworth filter used in the LP-MPPI algorithm (top row) and the control trajectories generated by it (bottom row) for a range of cutoff frequencies $\cutoff$ and filter orders $\order$.}
    \label{fig:butter}
    \vspace{-3mm}
\end{figure}

Our proposed algorithm, LP-MPPI, does not introduce any additional parameters beyond those of a low-pass filter. Therefore, in our case, the parameters of the LP-MPPI are the parameters of the Butterworth filter itself -- cutoff frequency $\cutoff$ and order $\order$. The cutoff frequency controls the width of the passband, which affects the range of frequencies with the highest power density in the drawn control signal perturbations. A lower $\cutoff$ results in a greater bias at lower frequencies and an increased use of low-frequency signals. In turn, a higher $\cutoff$ reduces the bias on low frequencies and allows searching in a wider range of signals.
The second parameter of the Butterworth filter is its order $\order$, which controls the slope of the power density spectrum in the stopband -- roll-off, i.e., how much the attenuation of the signal components grows with the growth of the frequency. The higher the order, the lesser the exploration outside the passband, and the greater the bias towards low frequencies.
To better visualize these dependencies, in Figure~\ref{fig:butter} we present the impact of the filter cutoff frequency $\cutoff$ and its order $\order$ on the magnitude of the frequency response and on the signals in the time domain.

Last but not least, one of the benefits of the proposed approach is the interpretability of the cutoff frequency $\cutoff$ parameter, as it enables intuitive tuning. One of the natural approaches to choosing this parameter is to set it around the natural cutoff frequency of the system to be controlled with LP-MPPI, as inducing higher frequencies requires a significant effort to overcome the damping characteristics of the system itself. Moreover, one can often heuristically estimate the highest expected frequency of the system states necessary to maximize the reward or obtain the desired behavior. Furthermore, in some cases, the interpretability of the cutoff frequency may help regularize the system's behavior, e.g., preventing exploitation of the carelessly designed reward function.



To sum up, the proposed LP-MPPI approach filters out the high-frequency components from the evaluated control signals within the MPPI loop, significantly reducing the high-frequency components in the obtained controls. This reduction is particularly important for robotic systems, as it decreases their wear and biases the search for high-performance control signals towards a more promising area (consider the plausibility of finding a performant control signal in the upper left and bottom right corners of the bottom of Figure~\ref{fig:butter}). Finally, our approach enables one to effectively control the frequency spectrum of the control signal perturbations, while providing intuitive tuning.




\section{Experiments}

\newcommand{\heightenvs}{16.5mm}
\begin{figure}[t]
    \centering
    \setlength\tabcolsep{0.5mm}
    \begin{tabular}{ccc}
         \scriptsize \textbf{Hopper-v5} & \scriptsize \textbf{Ant-v5} & \scriptsize \textbf{HalfCheetah-v5}\\
         \includegraphics[height=\heightenvs]{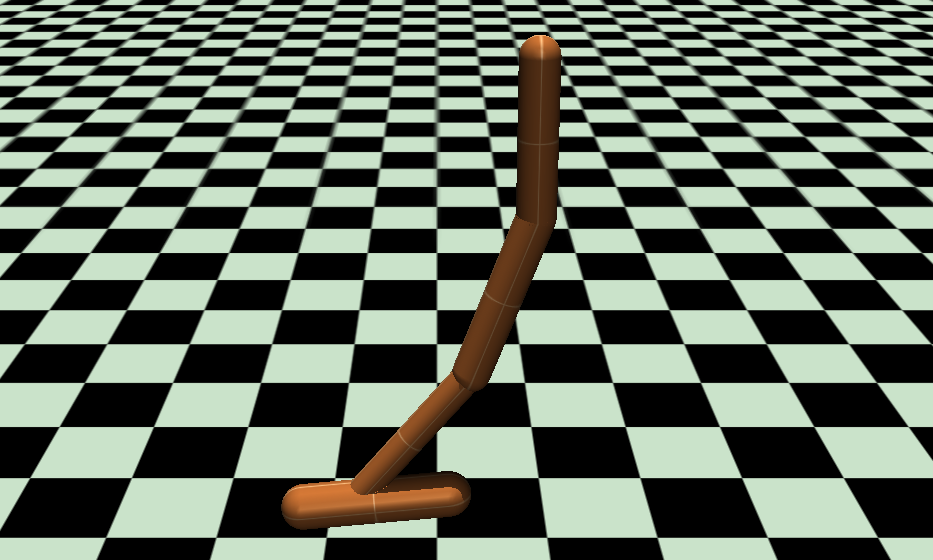} &
         \includegraphics[height=\heightenvs]{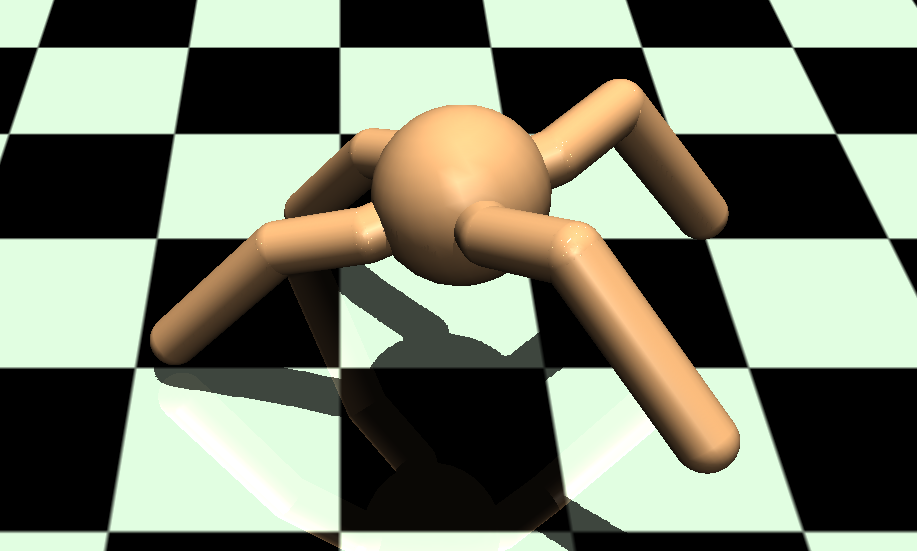} &
         \includegraphics[height=\heightenvs]{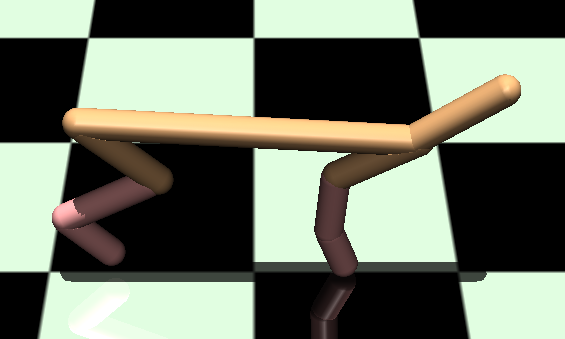}\\
         \scriptsize \textbf{Unitree Go2} & \scriptsize \textbf{MAB Silver Badger} & \scriptsize \textbf{F1TENTH racing} \\
         \includegraphics[height=\heightenvs]{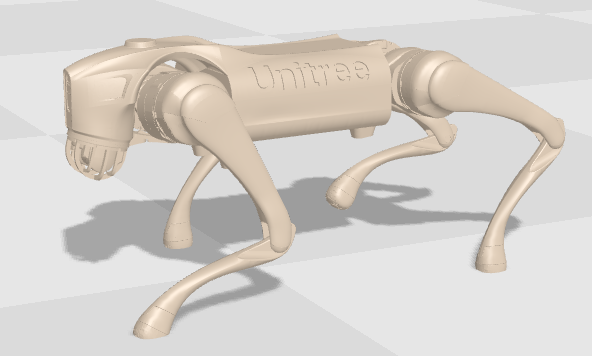} &
         \includegraphics[height=\heightenvs]{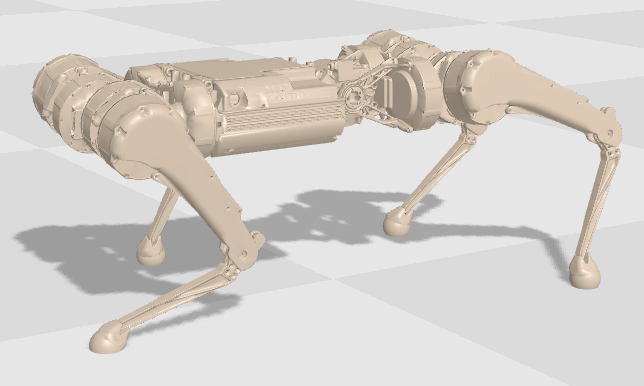} &
         \includegraphics[height=\heightenvs]{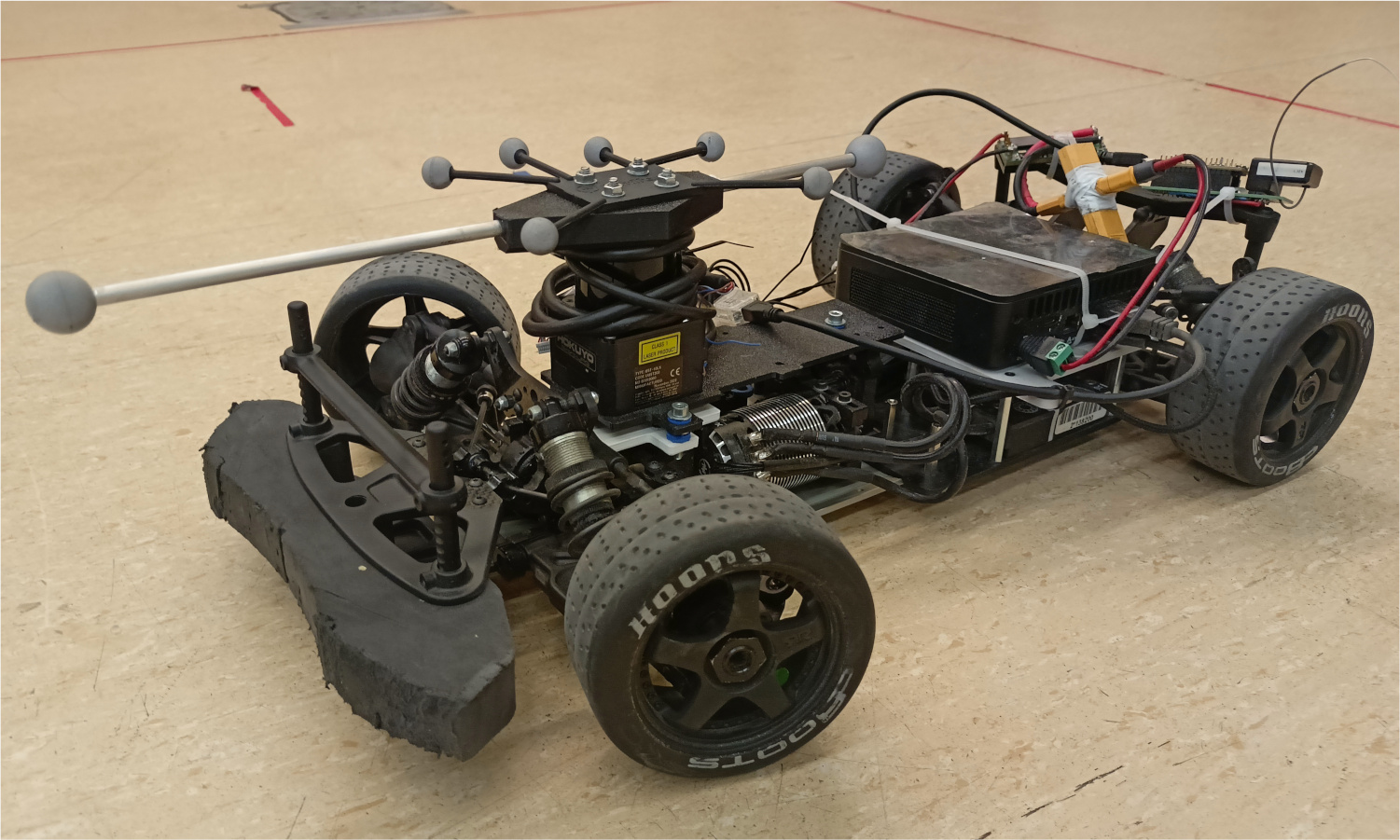}\\
    \end{tabular}
    \vspace{-2mm}
    \caption{Systems used in the experimental evaluation.}
    \label{fig:envs}
    \vspace{-3mm}
\end{figure}

\subsection{Gymnasium environments}
Our first experiment is an optimal control task in 3 high-dimensional Gymnasium environments~\cite{gymnasium}: \textit{Hopper-v5}, \textit{Ant-v5}, and \textit{HalfCheetah-v5} (see Figure~\ref{fig:envs}).
The goal of this experiment is to evaluate the efficiency of the proposed approach and relate it to the State-of-the-Art MPPI-based control algorithms.
We compared our proposed method with the following baselines:
\begin{itemize}
    \item MPPI -- the original MPPI algorithm~\cite{mppi},
    \item SCP-MPPI -- Spline-interpolated MPPI~\cite{splinemppi},
    \item SMPPI -- smoothed MPPI by control space lifting~\cite{smppi},
    \item ColoredMPPI -- MPPI with colored noise~\cite{colored}.
\end{itemize}

In this experiment, we assumed the perfect knowledge of the system dynamics, i.e., we used the same simulated environments for MPPI rollouts and evaluation. As a cost function, we used the original rewards from the Gymnasium environments, taken with a minus sign. To ensure a fair comparison, we tuned the parameters of each algorithm using Optuna~\cite{optuna} and 100 trials. In addition, we averaged the results over 100 episodes to reduce the impact of randomness.

\newcommand{\height}{16.2mm}
\newcommand{\heightt}{18.2mm}
\begin{figure*}[t]
    \centering
    \setlength\tabcolsep{0.5mm}
    \begin{tabular}{p{3mm}cp{3mm}cccc}
         & & & \multicolumn{4}{c}{\footnotesize \textbf{Improvement of LP-MPPI over}}\\
         & \scriptsize \textbf{LP-MPPI rewards (ours)} & & \scriptsize \textbf{MPPI \cite{mppi}} & \scriptsize \textbf{ColoredMPPI \cite{colored}} & \scriptsize \textbf{SMPPI \cite{smppi}} & \scriptsize \textbf{SCP-MPPI \cite{splinemppi}} \\
         \rotatebox[origin=l]{90}{\parbox[c]{1.5cm}{\centering \scriptsize \quad \textbf{Hopper-v5}}} &
         \includegraphics[height=\heightt]{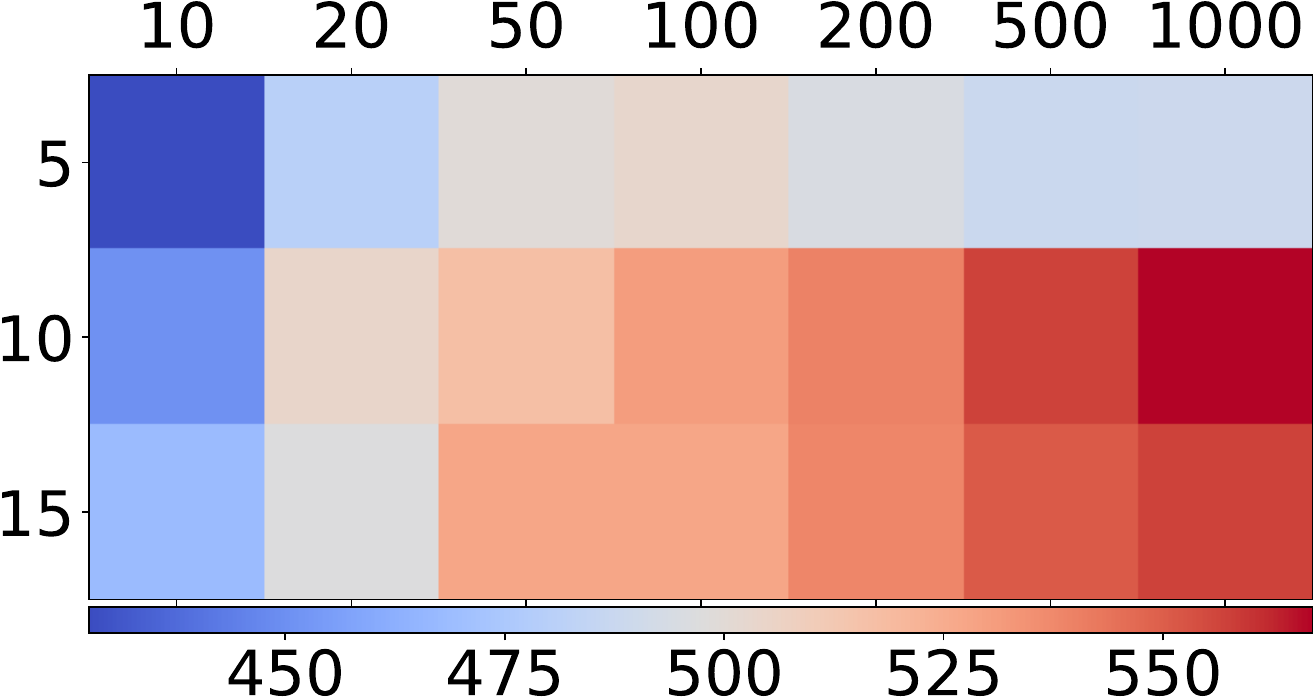} & &
         \includegraphics[height=\heightt]{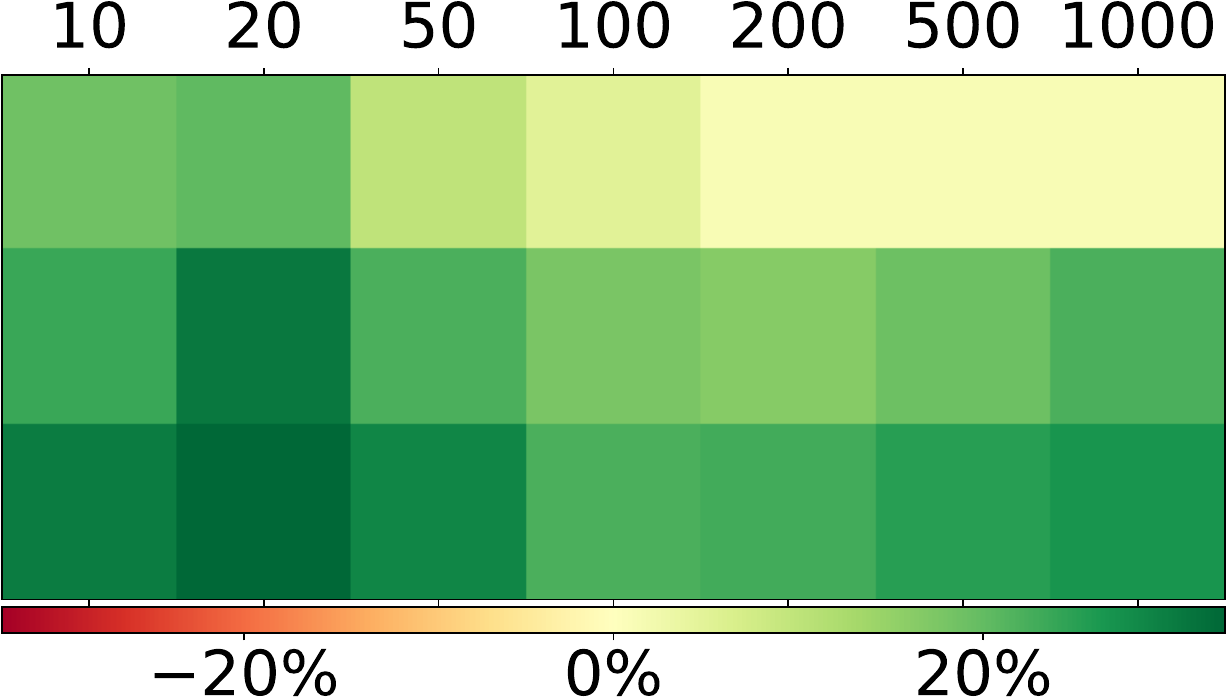} &
         \includegraphics[height=\heightt]{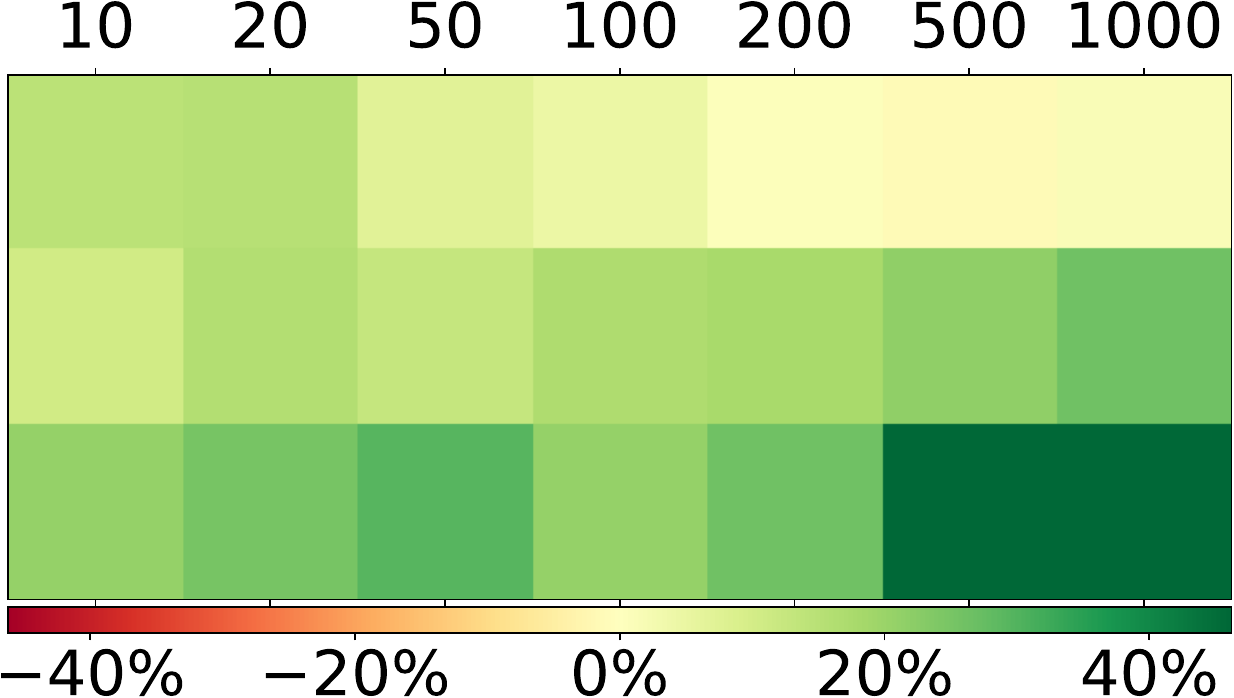} &
         \includegraphics[height=\heightt]{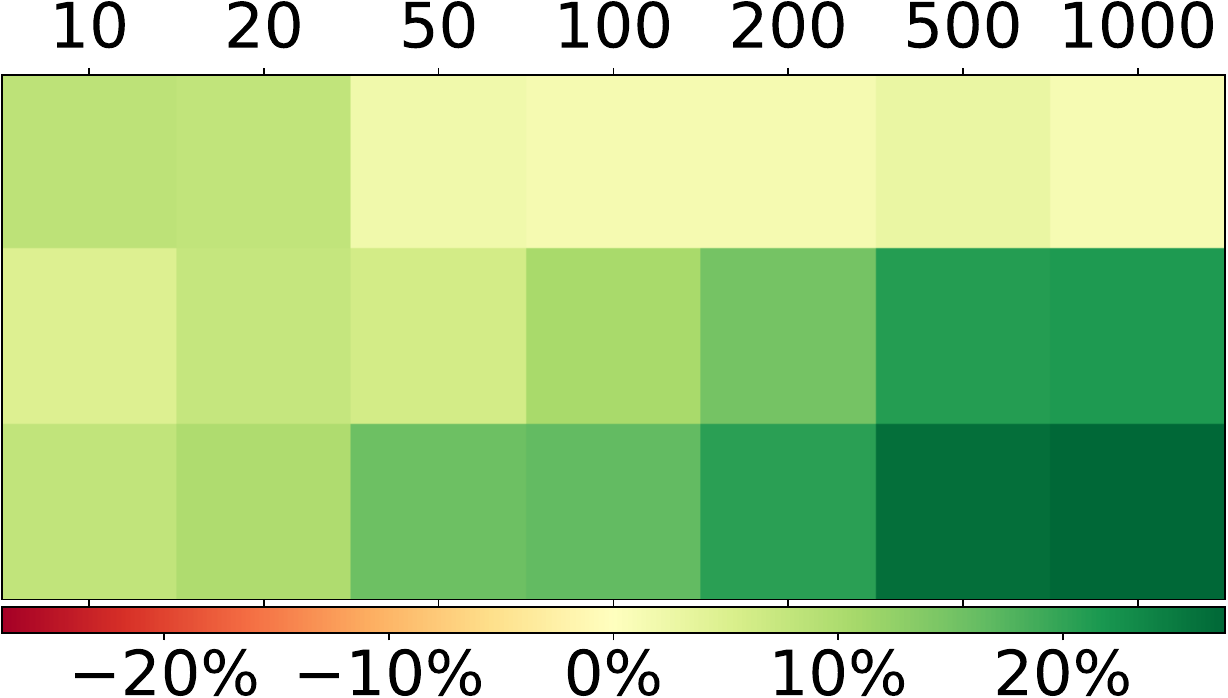} &
         \includegraphics[height=\heightt]{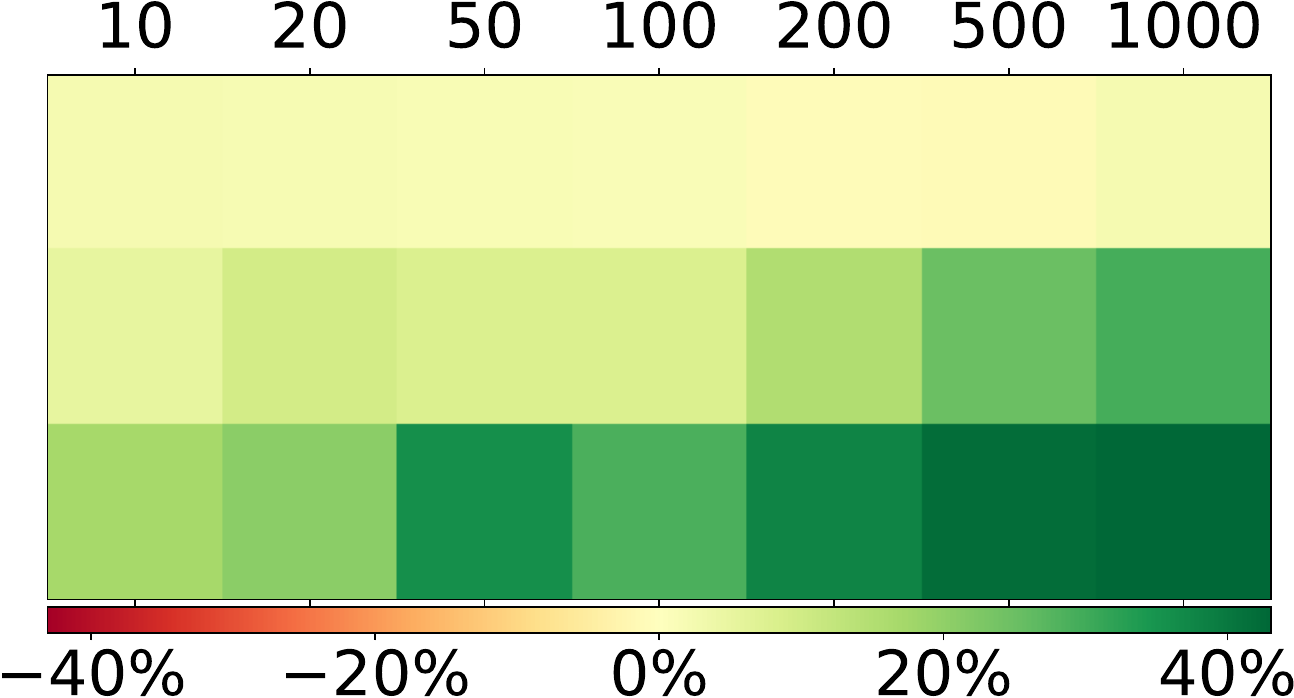}\\
         \rotatebox[origin=l]{90}{\parbox[c]{1.5cm}{\centering \scriptsize \quad \textbf{Ant-v5}}} &
         \includegraphics[height=\height]{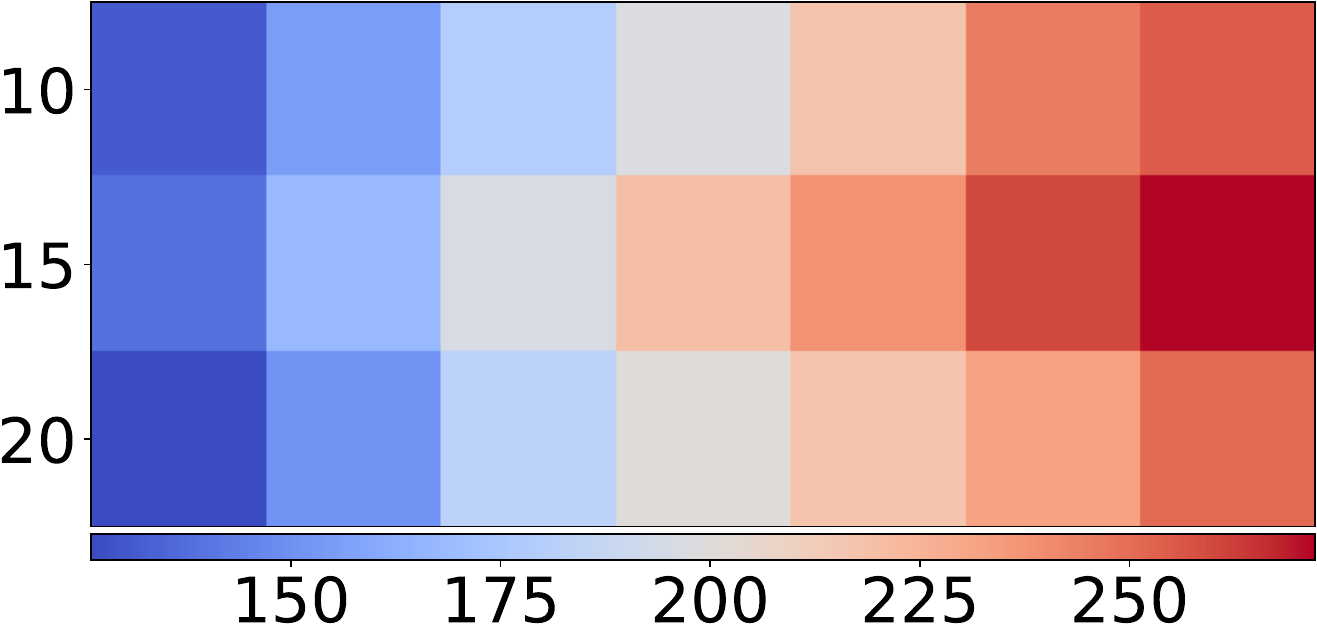} & &
         \includegraphics[height=\height]{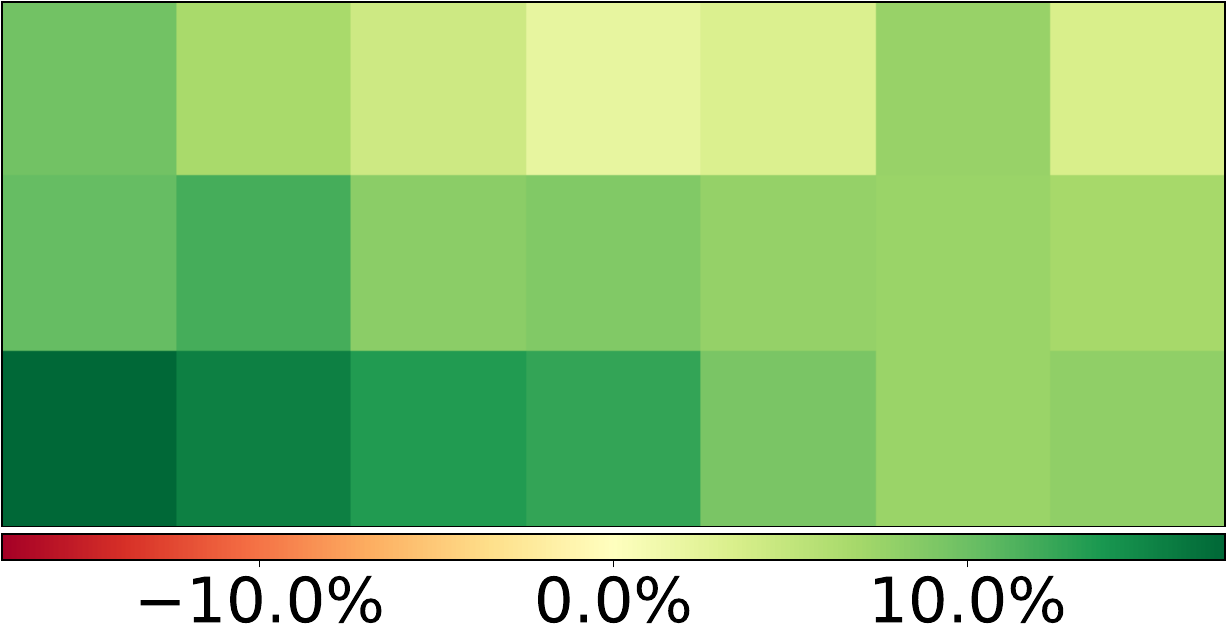} &
         \includegraphics[height=\height]{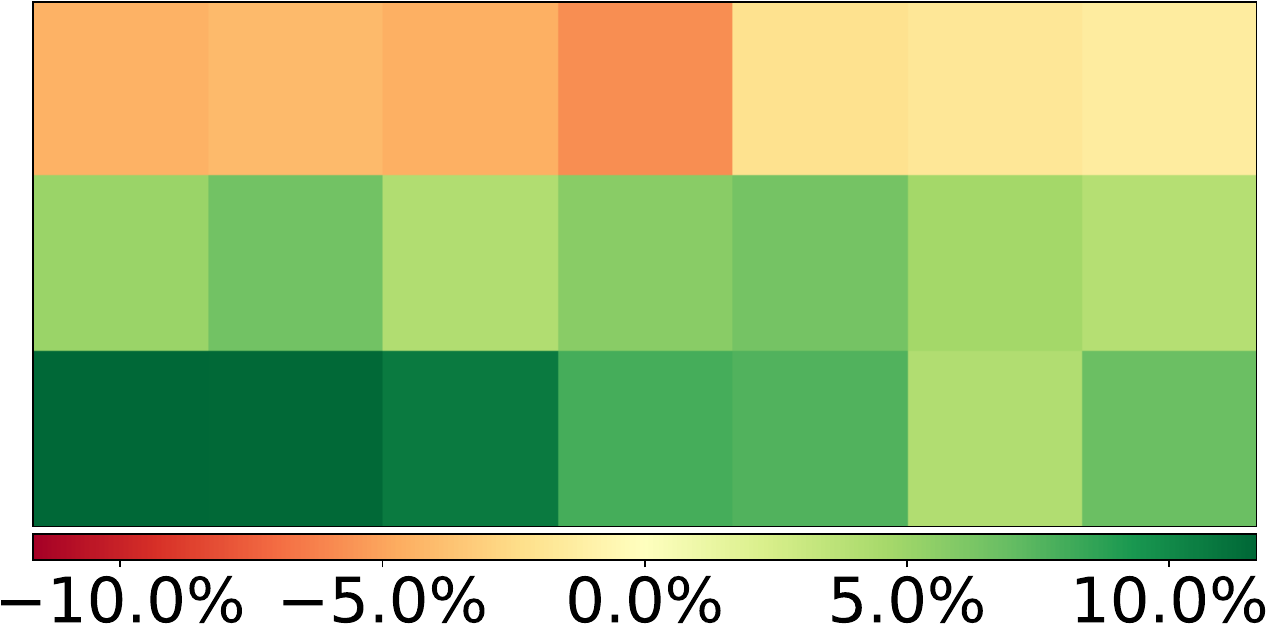} &
         \includegraphics[height=\height]{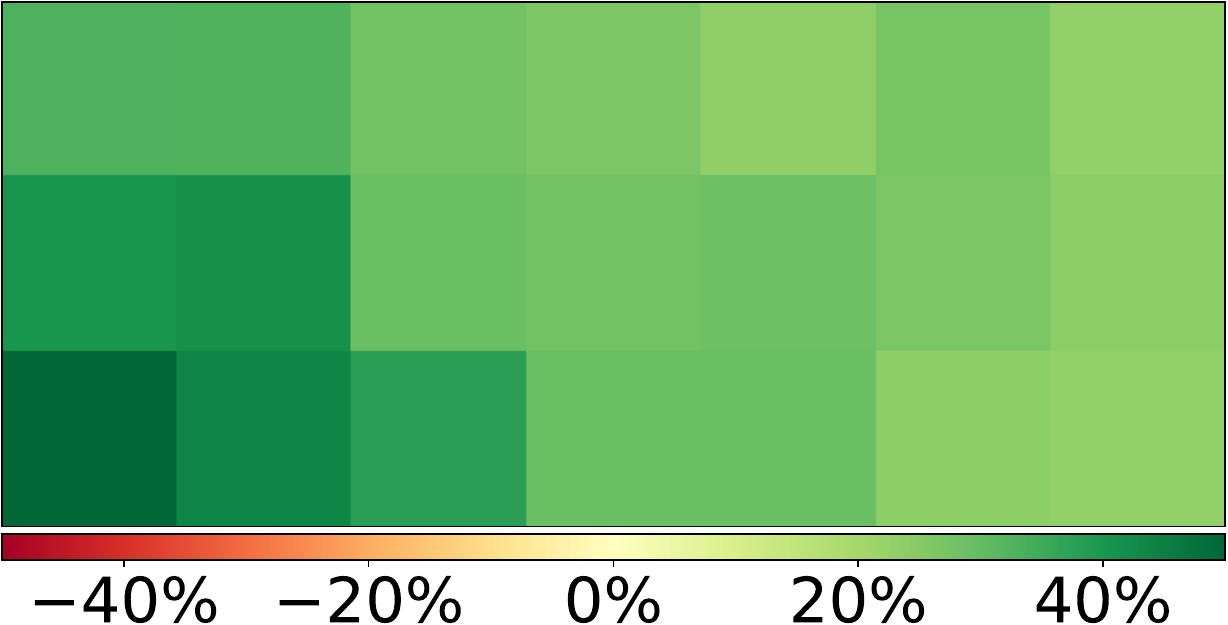} &
         \includegraphics[height=\height]{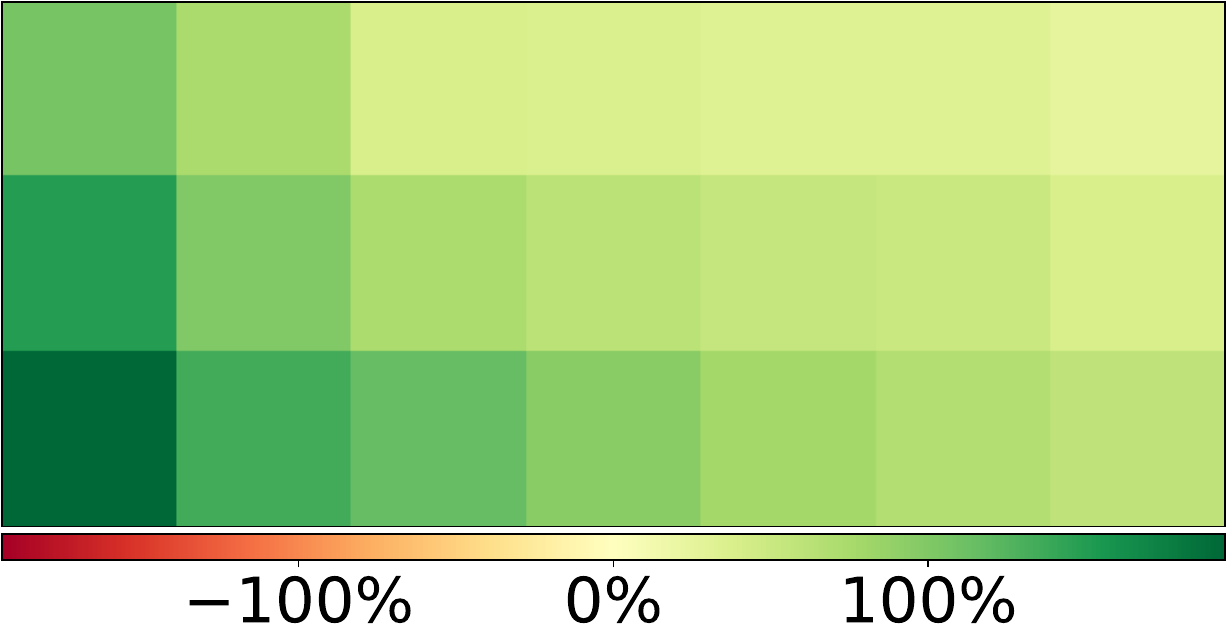}\\
         \rotatebox[origin=l]{90}{\parbox[l]{1.6cm}{ \hspace{-2mm} \scriptsize \textbf{HalfCheetah-v5}}} &
         \includegraphics[height=\height]{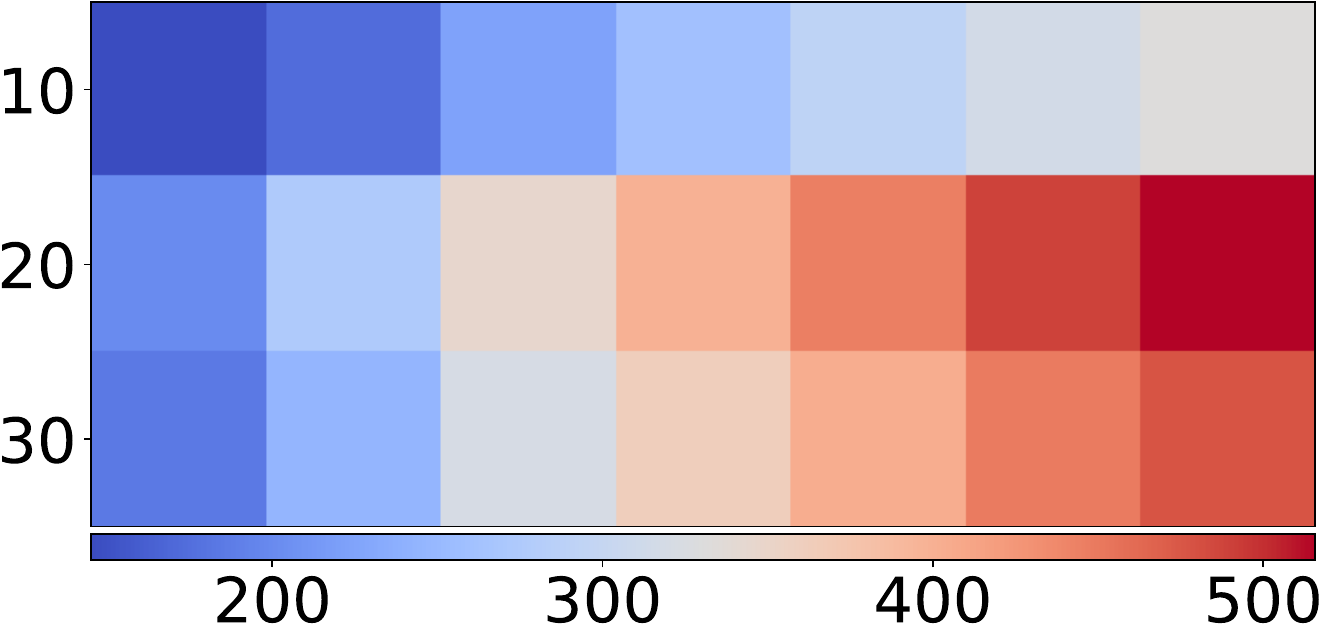} & &
         \includegraphics[height=\height]{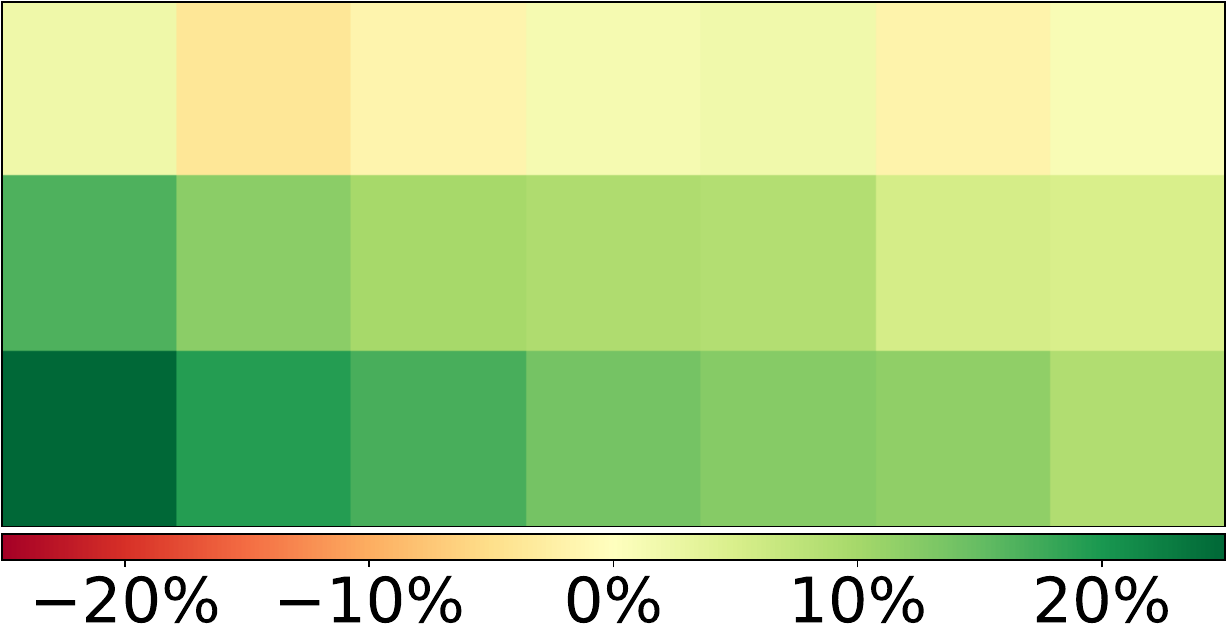} &
         \includegraphics[height=\height]{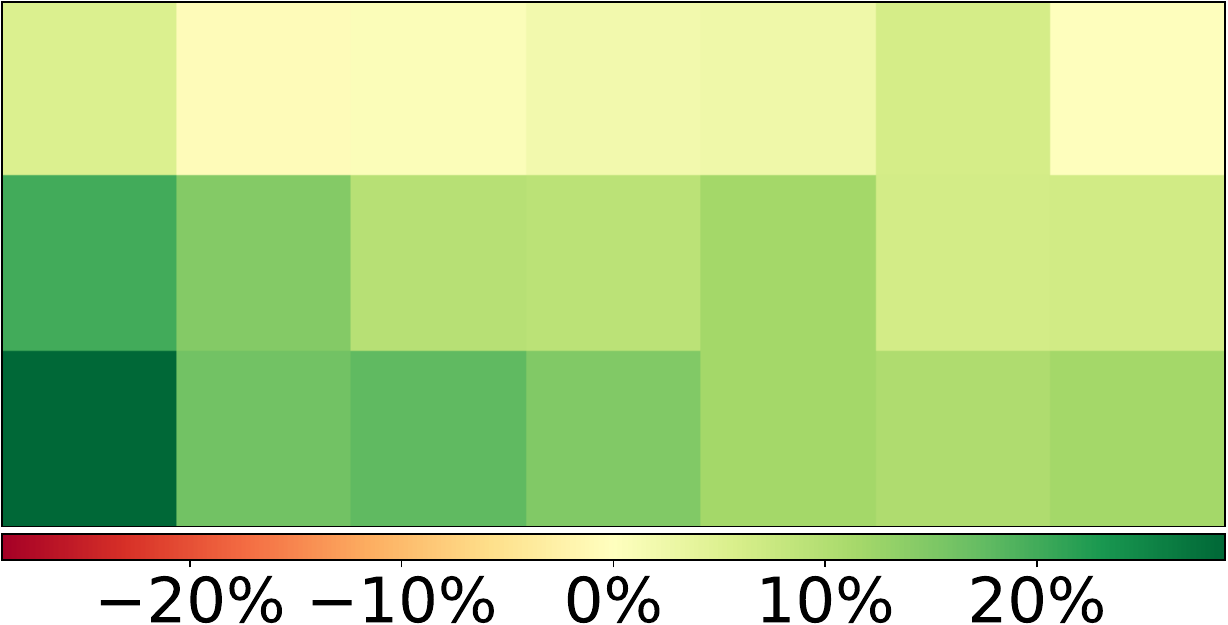} &
         \includegraphics[height=\height]{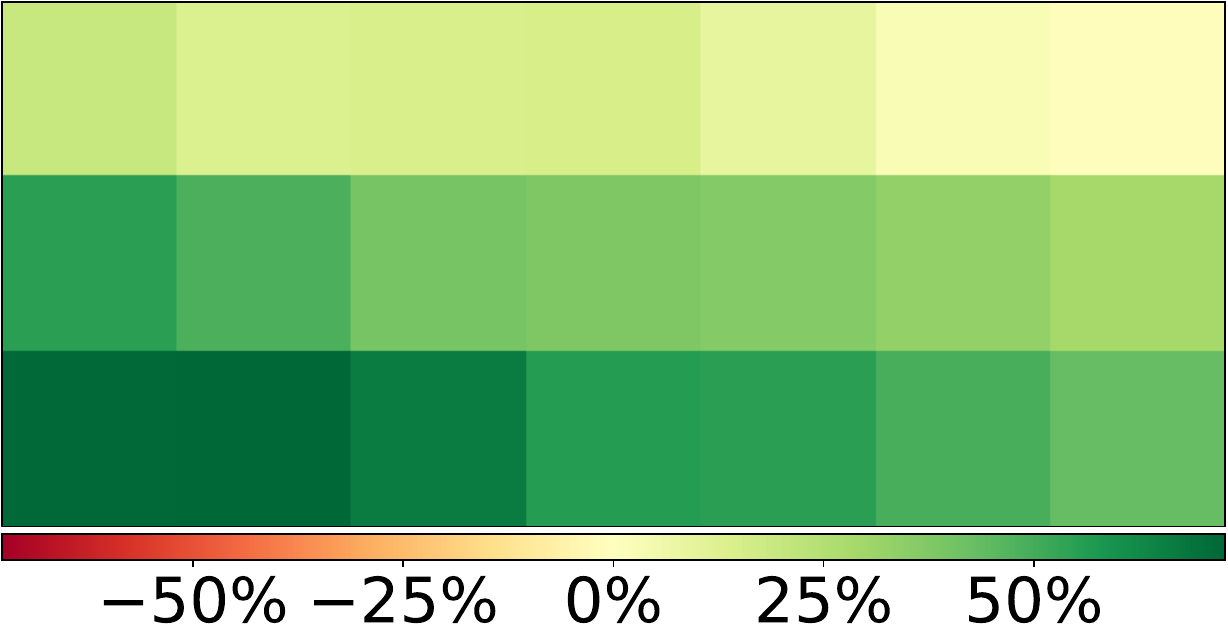} &
         \includegraphics[height=\height]{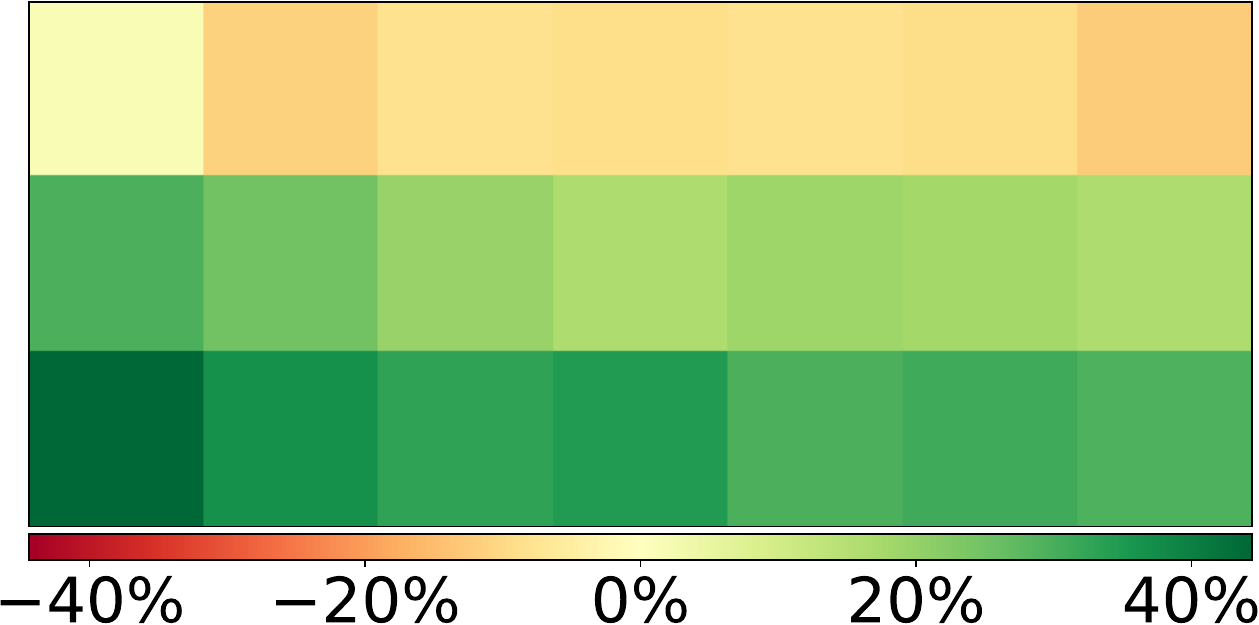}\\
    \end{tabular}
    \vspace{-2mm}
    \caption{Performance of the proposed LP-MPPI algorithm in Gymnasium environments (left), for a range of horizon lengths $H$ (y-axis) and numbers of rollouts $N$ (x-axis), and comparison to baselines (right). The green color represents the situation in which the LP-MPPI outperforms the baseline.}
    \label{fig:gym}
    \vspace{-4mm}
\end{figure*}

The results of our experiment are presented in Figure~\ref{fig:gym}. We evaluated each of the method, in each environment, on a matrix of the common MPPI parameters, i.e., horizon length $H$ (vertical axis), and number of rollouts $N$ (horizontal axis).
In the first column, one can see the rewards obtained by the LP-MPPI, while in the remaining columns the relative improvement of the LP-MPPI over the baselines. The superiority of the proposed approach can be seen by the dominance of the green color in the presented chart. In fact, except for very short horizons, LP-MPPI outperforms all baselines in all environments considered, no matter the number of rollouts. The scale of the improvement vary for different horizons and numbers of rollouts, but in general, the longer the horizon, the bigger the improvement. 
For \textit{Ant-v5} and \textit{HalfCheetah-v5} environments on can observe that the improvements are larger for a smaller number of evaluated rollouts, showing the increased sample efficiency introduced by low-pass filtering of samples, which may be critical in computationally restricted edge devices. Interestingly, the biggest improvements are observed w.r.t. SMPPI and SCP-MPPI approaches, on average 26\% and 36\%, which may be caused by the lack of fine-grained control due to action representation. The best performance among the baselines is observed for ColoredMPPI~\cite{colored}, which is still, on average, about 10\% less performant than the proposed LP-MPPI. In turn, if we consider the results obtained for the best pair of horizon length and number of rollouts, then LP-MPPI outperforms it on average by 8.46\%.

\begin{figure}[t]
    \centering
    \includegraphics[width=\linewidth]{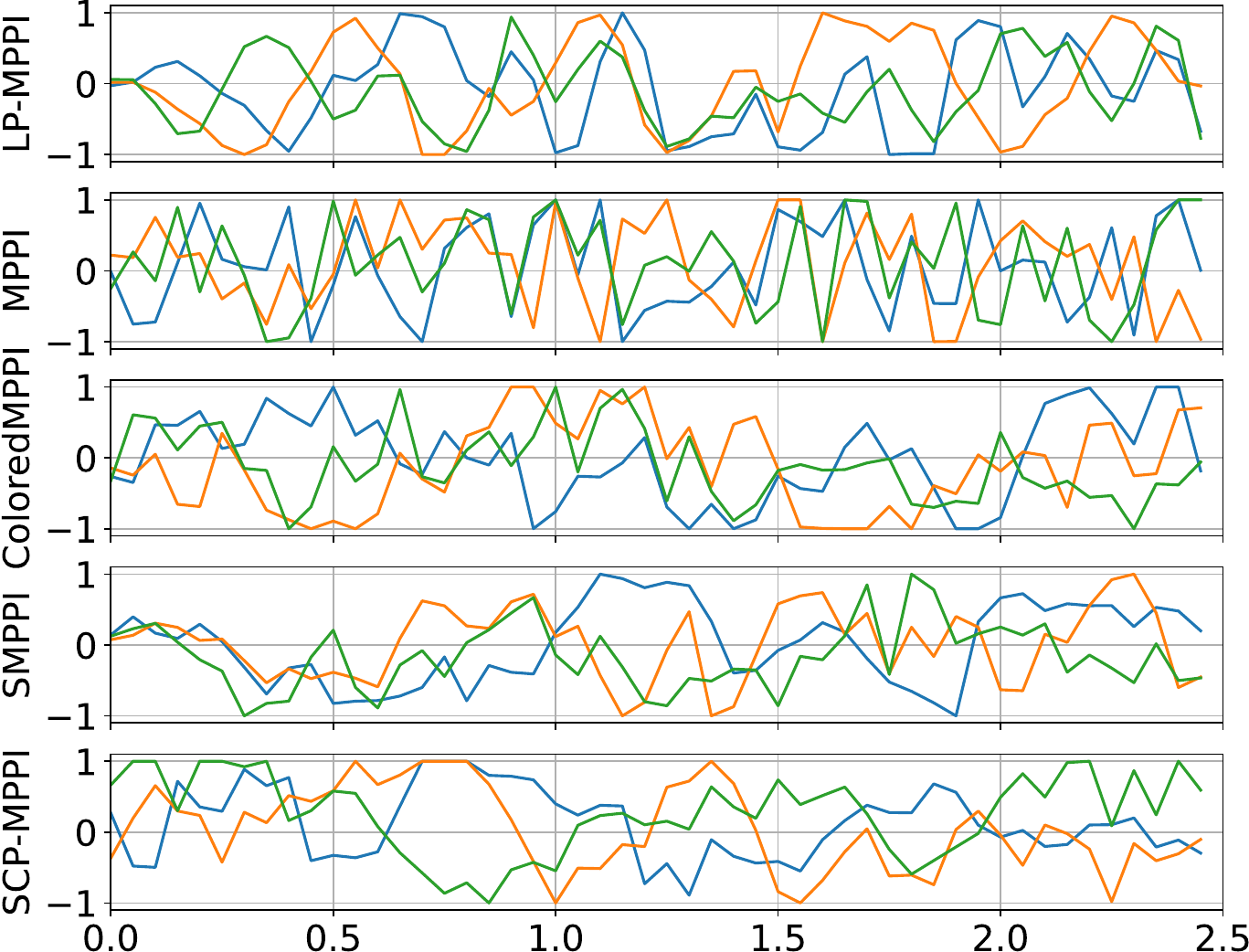}
    \vspace{-6mm}
    \caption{Sample applied control signals in the Ant-v5 environment for the considered MPPI-based control algorithms.}
    \label{fig:smoothness}
    \vspace{-6mm}
\end{figure}

Despite higher rewards, an important aspect of the control algorithm is the smoothness of the applied controls. To visualize the differences between the considered methods, we present the applied control trajectories for the Ant-v5 environment with $H=15$ and $N=100$ in Figure~\ref{fig:smoothness}. For fairness of comparison, each algorithm was used with the optimal set of parameters, w.r.t. cumulative reward, determined with Optuna~\cite{optuna} using 100 trials. 
One can see that the original MPPI algorithm generates undesirably sharp controls, while the remaining baselines produce significantly smoother controls, comparable to each other and to our proposed LP-MPPI.
To quantitatively assess smoothness, we report Mean Squared Second Derivative (MSSD) and Mean Savitzky-Golay Filter Deviation (MSGFD) in Table~\ref{tab:smoothness}.
The results indicate that our method attains very low MSSD and the smallest MSGFD, confirming that the control trajectories it generates are notably smooth.

\begin{table}[]
    \setlength\tabcolsep{1mm}
    \centering
    \caption{Comparison of control signals smoothness in Ant-v5 environment using the Mean Squared Second Derivative (MSSD) and Mean Savitzky-Golay Filter Deviation (MSGFD).}
    \vspace{-2mm}
    \begin{tabular}{c|ccccc}
         Metric & \textbf{LP-MPPI} & MPPI & ColoredMPPI & SMPPI & SCP-MPPI \\
         \hline
         MSSD & 0.409 & 2.144 & 0.805 & 0.402 & \textbf{0.350}\\
         MSGFD & \textbf{0.018} & 0.177 & 0.059 & 0.028 & 0.027\\
    \end{tabular}
    \label{tab:smoothness}
    \vspace{-4mm}
\end{table}

\subsection{Quadruped locomotion}
In the previous experiment, we showed that LP-MPPI outperforms the baselines for simplified abstract robots. Instead, in this one, we focus on the full-scale simulated quadrupeds with 12DoF. We consider a locomotion task with trot gait, imposed in the reward function to guide the search of the control signals, on two quadrupeds, i.e., Unitree Go2 and MAB Silver Badger (see Figure~\ref{fig:envs}).

Recently, it was shown that MPPI-based approaches can succeed in such complex control tasks by diffusion-style annealing of the standard deviation of the noise distribution and the use of spline interpolation, as it was done in the Dial-MPC approach~\cite{dialmpc}.
We would like to enhance the Dial-MPC with the proposed low-pass filtering (LP-Dial-MPC) to evaluate whether it applies to recent MPPI-based approaches and can improve their performance. Moreover, we would like to compare our method, in this setting, with the best-performing approach from the previous experiment -- ColoredMPPI~\cite{colored}. 
Thus, we extend the Dial-MPC with colored noise instead of the default white one.

In this experiment, we used the default settings of the Unitree Go2 trot experiment available in the code repository associated with the Dial-MPC paper~\cite{dialmpc}. The goal is to follow the desired longitudinal velocity of 1 m/s with the center of the robot trunk while maintaining its default orientation and height above the ground. 
In addition, a cost function that imposes a specific foot-height trajectory encourages the robot to follow a trot gait.
Moreover, we designed a very similar experiment for the MAB Silver Badger robot, with the same goals as for the Go2 but with additional cost terms regarding the energy consumption and the minimum required height of the calves, to encourage more \textit{natural} looking robot posture. In both experiments, we set the horizon $H=16$, number of rollouts $N=256$, $dt=20$ ms, temperature $\lambda = 0.05$, horizon and trajectory diffusion factors equal to $0.9$ and $0.5$, respectively, and the number of diffusion steps equal to $2$.

\begin{figure}[t]
    \centering
    \begin{tabular}{cc}
         \scriptsize \textbf{Unitree Go2} & \scriptsize \textbf{MAB Silver Badger} \\
         \includegraphics[height=39mm]{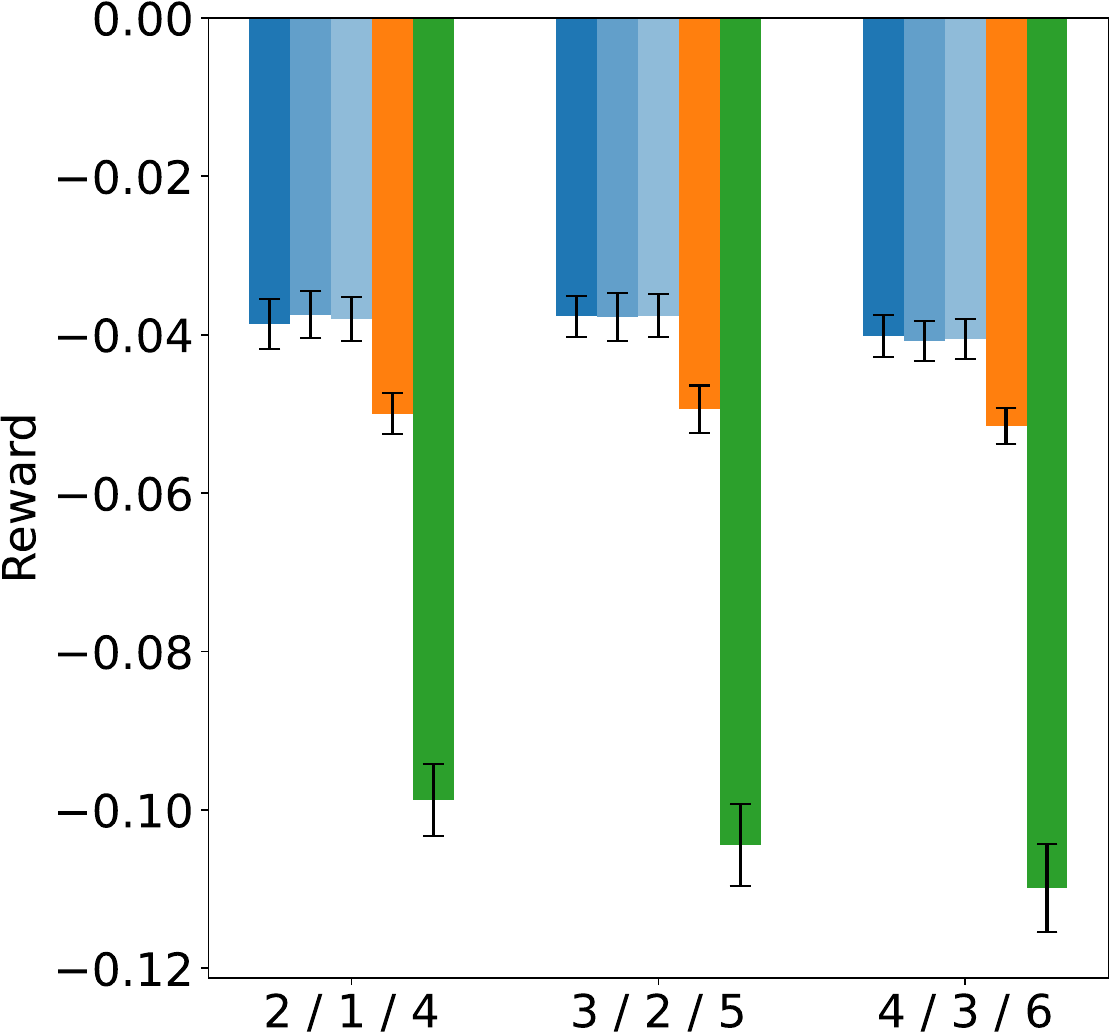} &
         \includegraphics[height=39mm]{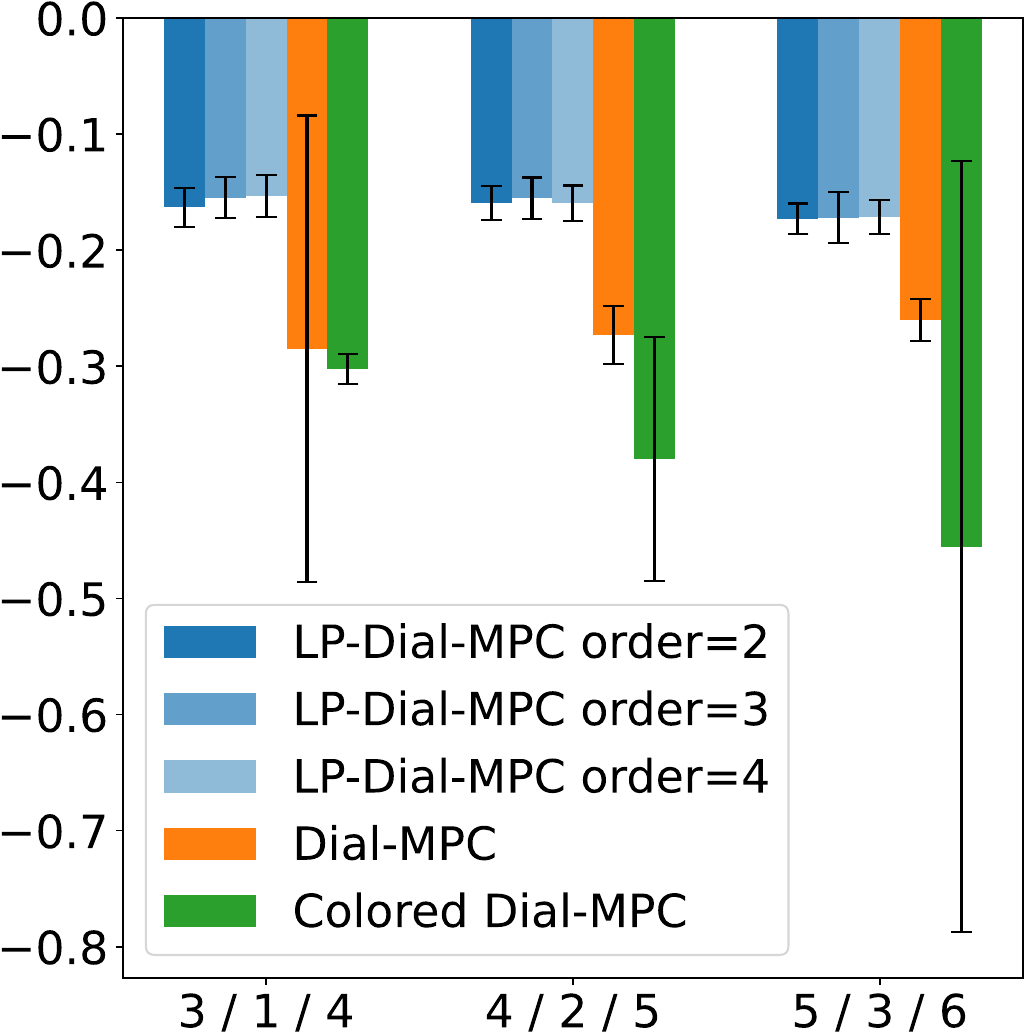}\\
         \multicolumn{2}{c}{\scriptsize $f_c$ [Hz] / $\beta$ [-] / $n_{cp}$ [-] }
    \end{tabular}
    \vspace{-2mm}
    \caption{Comparison of the Dial-MPC~\cite{dialmpc} with the proposed low-pass filtering and the baseline colored noise~\cite{colored} perturbations distribution in the task of simulated quadruped locomotion.}
    \label{fig:quadruped}
    \vspace{-3mm}
\end{figure}

In this experiment, to highlight the robustness of the proposed approach to the choice of its parameters, we do not perform the search with Optuna but instead report the performance for several intuitive parameter sets, i.e. cutoff frequency $\cutoff \in \{2, 3, 4\}$ for Unitree and $\cutoff \in \{3, 4, 5\}$ for MAB robot, and orders $\order \in \{2, 3, 4\}$. We compared their performance with the two above-mentioned baselines, for which we found the sets of the best parameters using Optuna. The results of this experiment can be found in Figure~\ref{fig:quadruped}. The proposed low-pass filtering approach implemented into the Dial-MPC framework consistently outperforms the default Dial-MPC, by 24\% and 41\% for the Unitree Go2 and MAB Silver Badger robots, respectively. We attribute these improvements to the more fine-grained control (higher number of decision variables) of the LP Dial-MPC and its ability to directly shape the frequency spectrum of the sampling distribution. In turn, the spectrum shaping capabilities of the colored noise are relatively limited and bias only the lowest frequencies, which results in significantly worse performance (about two times lower rewards than ours).

\subsection{Real-world F1TENTH racing}
In all previous experiments, we assumed that the models of the controlled systems are perfectly known and are used by the MPPI to search for the best control trajectories. In turn, in this task, we would like to use an analytical model of the F1TENTH car (dynamic single-track model~\cite{iros} with MF6.1 tire model~\cite{besselink2010improved}) and evaluate it under the real-world racing conditions (see Figure~\ref{fig:envs}). The goal of this task is to cover the highest possible distance around the track centerline in 30 s, on the \SI{14.2}{\metre} long \SI{1}{\metre} wide oval racetrack.
We defined the cost function by
\begin{align*}
    \cost = &-v_f + 100\log(1 + \exp(-100(T_w/2 - n))) \\
    & + 100 \max(\alpha - 0.3, 0) + 2 (\theta - T_\theta)^2, 
\end{align*}
where $v_f$ is the velocity along the centerline, $n$ is the distance to the centerline, $T_w$ is the track width, $\alpha$ is the slip angle, $\theta$ is the vehicle orientation, and $T_\theta$ is the orientation of the centerline. We set the horizon $H=30$, $dt=50$ ms, control frequency to \SI{30}{\hertz}, and evaluated the algorithms for both $N=10$ and $N=50$ rollouts. We have chosen the number of rollouts and the horizon to meet real-time requirements with a single core of Intel Core i5-12500H CPU, while achieving reasonable driving performance (see video attachment). The parameters of all methods were chosen in simulation using Optuna~\cite{optuna} with 50 trials.

\begin{figure}[t]
    \centering
    \setlength\tabcolsep{2mm}
    \begin{tabular}{cc}
         \scriptsize \textbf{N=10} & \scriptsize \textbf{N=50} \\
         \includegraphics[height=42mm]{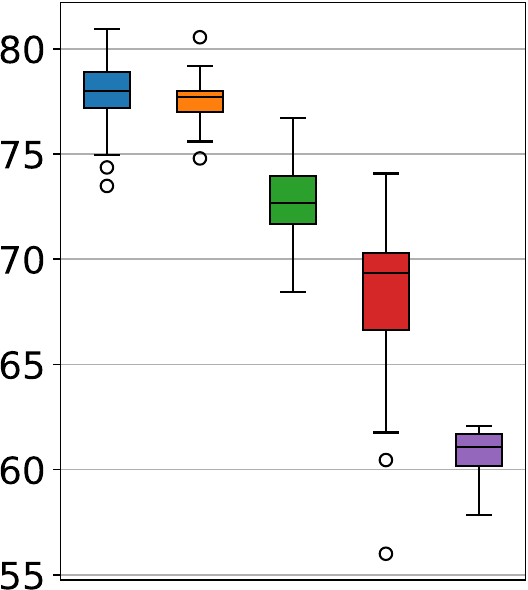} &
         \includegraphics[height=42mm]{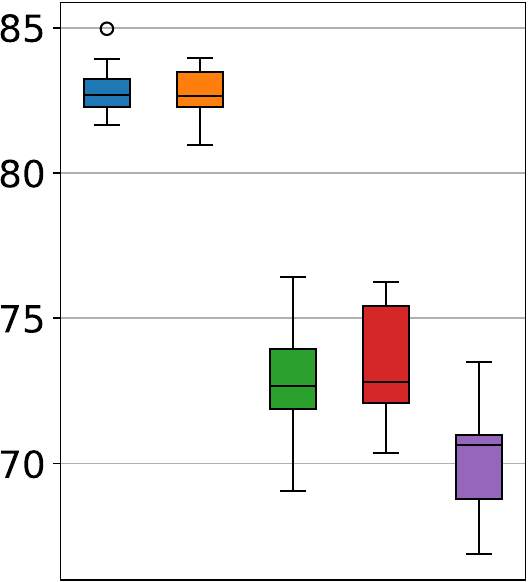}\\
    \end{tabular}
    \vspace{5mm}
    \includegraphics[width=\linewidth]{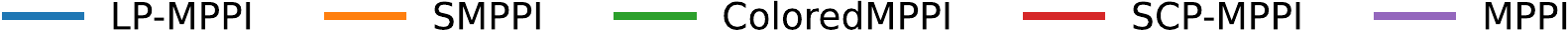}
    \caption{Distances [m] covered by the F1TENTH car in 30-second time trials using different MPPI algorithm variants.
    The proposed LP-MPPI achieves the highest median distances both for $N=10$ and $N=50$ rollouts.}
    \label{fig:f1tenth}
    \vspace{-3mm}
\end{figure}

In Figure~\ref{fig:f1tenth}, we present the distances covered by the proposed LP-MPPI and the other considered MPPI variants in 15 runs of \SI{30}{\second} each. One can see that in both considered setups, the proposed LP-MPPI approach significantly outperformed all baselines except SMPPI, which performed very close to the LP-MPPI for $N=50$ and a bit worse for $N=10$. Note that in the racing scenarios, even the \SI{30}{\centi\metre} of difference gained every \SI{30}{\second} of the race (the difference between medians of LP-MPPI and SMPPI) may be considered a notable gap.

\subsection{Computational overhead}
An important aspect of every control method is its computational efficiency. 
Therefore, we evaluate how much computational overhead the proposed method introduces relative to the nominal MPPI and how it relates to the other considered baselines.
To do so, we run all methods for 10 episodes in Ant-v5 (MuJoCo model, $H=15, N=100$) and F1TENTH (compiled analytical model, $H=30, N=10$) environments, using a single core of the Intel Core i5-12500H CPU. We compute the median of the control computation time and relate it to the one obtained by the MPPI.
In Table~\ref{tab:time}, we present the results of this experiment. One can see that all baselines introduce some notable computational overhead in the F1TENTH environment, since the compiled analytical model, which is responsible for most of the computations, is very fast. Note that the proposed method introduces the second smallest overhead of 2.4\%.
In turn, in the case of a relatively heavy dynamics model, e.g., Ant-v5 environment, we observe some counterintuitive results, like the decrease in the compute time for LP-MPPI and ColoredMPPI. We suppose that this may be caused by the variability in the timings, due to the use of a standard OS instead of a real-time one, or be an effect of filtering out the higher frequencies from the control signal, which may simplify the underlying physics simulation. In summary, a computationally intense dynamics evaluation causes the overhead introduced by the proposed method to be negligible.

\begin{table}[t]
    \centering
    \caption{Computational overhead relative to MPPI}
    \vspace{-2mm}
    \begin{tabular}{c|cccc}
         Environment & \textbf{LP-MPPI} & ColoredMPPI & SMPPI & SCP-MPPI \\
         \hline
         F1TENTH & +2.41\% & +5.12\% & +3.89\% & \textbf{+1.31\%}\\
         Ant-v5 & \textbf{-1.62\%} & -0.67\% & +3.04\% & +1.82\%\\
    \end{tabular}
    \label{tab:time}
    \vspace{-3mm}
\end{table}

\section{Conclusions}
In this work, we introduced Low-Pass Model Predictive Path Integral Control (LP-MPPI), a novel and easy-to-implement enhancement to MPPI that incorporates low-pass filtering into the sampling process. By directly shaping the frequency spectrum of control trajectory perturbations, LP-MPPI eliminates harmful high-frequency noise and improves the efficiency of searching for optimal control trajectories.
Unlike existing smoothing techniques or colored noise sampling, our approach offers intuitive fine-grained control over the optimal control search in the frequency domain, making it highly adaptable to various robotic systems.

Through extensive simulation and real-world experiments, we demonstrated the superiority of LP-MPPI over state-of-the-art MPPI-based methods in a variety of tasks, including simulated legged locomotion and real-world F1TENTH autonomous racing. Our results show that LP-MPPI consistently outperforms state-of-the-art methods by 10\% in Gymnasium environments, 32\% in simulated quadruped locomotion, and by \SI{0.115}{\second} in a \SI{30}{\second} long F1TENTH autonomous time trial. In addition, it significantly reduces the chattering of the control signal, leading to smoother and more reliable actuation.
Moreover, LP-MPPI maintains computational efficiency, introducing only a negligible overhead compared to standard MPPI, making it practical for real-time applications.

To sum up, LP-MPPI represents a simple yet powerful modification to MPPI, making it an attractive option for real-time robotic control tasks requiring both high-performance trajectory optimization and smooth, actuator-friendly control signals. Future work will explore adaptive filtering techniques to dynamically adjust the sampling distribution based on task demands and further integrate LP-MPPI with learning-based sampling strategies for improved adaptability.

\bibliographystyle{IEEEtran}
\bibliography{references}

\end{document}